\documentclass[10pt,a4paper]{amsart}

\usepackage{amsthm,amssymb,amsmath,epic,eepic,float}

\def\barx{\bar{x}}
\def\bary{\bar{y}}

\def\leq{\leqslant}
\def\geq{\geqslant}

\def\zi{Z_{N\! \times \! N}}
\def\zs{Z_{n\! \times \! n}}

\newtheorem{myLemma}{Lemma}
\newcommand{\myProof}{\noindent \emph{Proof}.\ }

\newcommand{\cA}{\mathcal A}

\newcommand{\cC}{\mathcal C}

\newcommand{\cM}{\mathcal M}
\newcommand{\cN}{\mathcal N}
\newcommand{\cO}{\mathcal O}
\newcommand{\cP}{\mathcal P}

\def\ll{ \left\lgroup}
\def\rr{\right\rgroup}

\newcommand{\1}{{\bf 1.}}
\newcommand{\2}{{\bf 2.}}
\newcommand{\3}{{\bf 3.}}
\newcommand{\4}{{\bf 4.}}

\def\union{\mathop{\bigcup}}

\def\det{\operatorname{det}}

\begin{document}
\title[Partial domain wall partition functions]
{Partial domain wall partition functions}
\author[O Foda and M Wheeler]
{O Foda $\!\! ^1$ and M Wheeler $\!\! ^2$}
\address{
$^1$ Department of Mathematics and Statistics,
University of Melbourne,
Parkville, Victoria 3010, Australia
}
\address{
$^2$ Laboratoire de Physique Th\'eorique et Hautes Energies,
CNRS UMR 7589 and Universit\'e Pierre et Marie Curie - Paris 6,
4 place Jussieu, 75252 Paris cedex 05, France
}

\email{omar.foda@unimelb.edu.au, mwheeler@lpthe.jussieu.fr}
\begin{abstract}
We consider six-vertex model configurations on an $(n \times N)$ 
lattice, $n$ $\leq$ $N$, that satisfy a variation on domain wall 
boundary conditions that we define and call {\it partial domain 
wall boundary conditions}.
We obtain two expressions for the corresponding 
{\it partial domain wall partition function}, as an 
$(N\! \times \! N)$-determinant and as an 
$(n\! \times \! n)$-determinant. 
The latter was first obtained by I Kostov.
We show that the two determinants are equal, as expected from 
the fact that they are partition functions of the same object, 
that each is a discrete KP $\tau$-function, 
and, recalling that these determinants represent tree-level 
structure constants in $\cN = 4$ SYM, we show 
that introducing 1-loop corrections, as proposed by 
N Gromov and P Vieira, preserves the determinant structure.
\end{abstract}
\maketitle
\setcounter{section}{-1}

%%%SEC00%%%
\section{Introduction}
\label{section-0}

The discovery of classical and quantum integrable structures on both 
sides of the anti-de Sitter/conformal field theory correspondence, 
AdS/CFT, in the late 1990's and early 2000's, culminating in 
\cite{bena.polchinski.roiban, minahan.zarembo} 
and in the intensive rapid developments that followed these seminal 
works, has been beneficial to all subjects involved 
\footnote{\ For a comprehensive introduction to applications 
of classical and quantum integrability in gauge and string theories, 
we 
refer the reader to \cite{beisert.review} and references therein.}.
On the one hand, integrability is widely considered to be a viable 
approach to proving the AdS/CFT correspondence. On the other, ideas 
and insights from AdS/CFT will continue to enrich the subject 
of integrability.  

\subsection{Partial domain wall boundary conditions}
The purpose of this note is to study six-vertex model configurations 
on a rectangular lattice with $n$ horizontal and $N$ vertical lines, 
$n \leq N$, as in Figure {\bf \ref{lattice}}, that satisfy 
$(n \! \times \! N)$ 
{\it partial domain wall boundary conditions}, pDWBC's. 
In our conventions
\footnote{\ Our conventions include choosing $n \leq N$, 
all $2n$ arrows on the left and right boundaries and $(N-n)$ arrows 
on the upper and lower boundaries point inwards, while all other 
arrows on the upper and lower boundaries point outwards. All these 
choices could have been reversed.}, these are defined as follows.

\begin{enumerate}
\item[\1] All arrows on the left and right boundaries point inwards,

\item[\2] $n_u$ ($n_l$) arrows on the upper (lower) boundary, 
such that $n_u+n_l = N-n$, 
also point inwards, 

\item[\3] The remaining $n+N$ arrows on the upper and lower boundaries 
point outwards,

\item[\4] The locations of the inward-pointing arrows on the upper 
and lower boundaries, with $n_u$ and $n_l$ fixed, are summed over. 
\end{enumerate} 

\noindent The corresponding partition function is 
an $(n\! \times \! N)$ {\it partial domain wall partition function}, 
pDWPF. 
For $n = N$, and $n_u = n_l = 0$, we recover Korepin's 
domain wall boundary conditions, DWBC's \cite{korepin.paper}, 
and the pDWPF reduces to Izergin's domain wall partition 
function, DWPF \cite{izergin.paper, korepin.book}. 

\subsection{A brief history of partial domain wall configurations}  
The configurations considered in this work were first 
introduced in the work of Bogoliubov, Pronko and Zvonarev in their 
study of boundary correlation functions in the presence of DWBC's 
\cite{bogoliubov.pronko.zvonarev}, and subsequent works 
\cite{foda.preston, colomo.pronko.1, colomo.pronko.2, colomo.pronko.3}. 
However, in these works, they were camouflaged by the fact that 
they were paired with complementary configurations to produce 
$(N \! \times \! N)$ configurations with conventional DWBC's, and 
that only a subset of the positions that the inverted $n_u$ ($n_l$) 
arrows on the upper (lower) boundaries can take were included in the 
statistical sum, as one does in computations of boundary correlation 
functions. 

In \cite{E1, E2}, Escobedo, Gromov, Sever and Vieira studied 3-point 
functions of three gauge-invariant single-trace length-$N_i$ 
operators, $\cO_i(x_i)$, $i \in \{1, 2, 3\}$, that are composed of 
elementary scalars in $SU(2)$ subsectors of $\mathcal{N} \! = \! 4$ 
supersymmetric Yang-Mills theory
\footnote{\  In \cite{E1, E2}, $\{\cO_i\}$, $i \in \{1, 2, 3\}$ are 
chosen such that their lengths $L_i$ satisfy {\it non-extremal 
length} conditions, $L_i < L_j + L_k$ for any distinct $\{i, j, k\}$, 
and further, they are characterized by rapidity variables $\{x_i\}$,
such that they are non-BPS ($\{x_i\}$ has finitely many elements 
that are finite, rather than infinite), and have well-defined 
conformal dimensions (the elements of $\{x_i\}$ satisfy Bethe 
equations).}. 
Using the connection with quantum integrable models, Escobedo 
{\it et al.} obtained a sum expression for the studied structure 
constants. In \cite{GSV}, Gromov, Sever and Vieira considered the 
same structure constants as in \cite{E1, E2} in the special case 
where $\cO_1$ and $\cO_3$, in the conventions of \cite{GSV}, are BPS 
operators. That is, the rapidity variables $\{x_1\}$ and $\{x_3\}$ 
that characterize these two operators are taken to infinity. 
In this special case, the sum expression of \cite{E1} simplifies 
to a sum expression for a quantity that they refer to as 
${\cA} \{x_2\}$. 

In \cite{F}, the sum expression of \cite{E1, E2} was evaluated in 
determinant form. This determinant is essentially Slavnov's 
determinant expression for the scalar product of a Bethe eigenstate 
and a generic state in a periodic spin-$\frac{1}{2}$ XXX chain. In 
vertex model terms, Slavnov's determinant is the partition function 
of rational six-vertex configurations with $2n$ horizontal lines 
(auxiliary spaces), $N$ vertical lines (quantum spaces), and boundary 
conditions that are specified in \cite{wheeler, F}. 
The limit used in \cite{GSV} to produce $\mathcal{A}\{x_2\}$ is achieved 
in six-vertex model terms by simply deleting the $n$ horizontal lines 
that represent the Bethe eigenstate, as explained in the sequel. The 
resulting configurations are the partial domain wall configurations 
discussed in this note.  

\subsection{Outline of contents}
In Section {\bf \ref{section-1}}, we recall basic definitions 
related to the six-vertex model. 
In Section {\bf \ref{section-2}}, 
we start from $(N\! \times \! N)$ domain wall configurations and delete 
$N-n$ horizontal lines to obtain $(n\! \times \! N)$ partial domain wall 
configurations. For simplicity we consider the case $n_u = N - n$
and $n_l = 0$. Once this case is understood, the general case is 
straightforward to obtain. The corresponding $(n\! \times \! N)$ pDWPF
is obtained starting from Izergin\rq{}s $(N\! \times \! N)$ determinant 
expression for the $(N \! \times \! N)$ DWPF of the initial domain wall 
configuration, taking the rapidity variables of the lines that are 
deleted to infinity, and normalizing appropriately to obtain the pDWPF 
$Z_{N \! \times \! N}$ in $(N \! \times \! N)$ determinant form.

In Section {\bf \ref{section-3}}, we start from the $(2n\! \times \! N)$ 
configurations that describe the scalar product of an $n$-magnon Bethe 
eigenstate and an $n$-magnon generic state, on an $N$-site periodic 
spin-$\frac{1}{2}$ chain. 
We delete the $n$ horizontal lines that describe the Bethe eigenstate 
to obtain $(n \! \times \! N)$ partial domain wall configurations with 
$n_u = N - n$ and $n_l = 0$.
The corresponding $(n \! \times \! N)$ pDWPF is obtained starting from 
Slavnov\rq{}s $(n \! \times \! n)$ determinant expression for the scalar 
product, taking the rapidity variables of the lines that we deleted 
(which are the Bethe roots) to infinity, and normalizing appropriately 
to obtain the pDWPF 
$Z_{n \! \times  \! n}$ in $(n \! \times  \! n)$ determinant form.
$Z_{n \! \times  \! n}$ was first derived by Kostov 
\cite{kostov.private.communication, kostov.short.paper, 
                                    kostov.long.paper}.
Expanding $Z_{n \! \times \! n}$, one obtains the sum expression of 
Gromov {\it et al.} \cite{E1}. Starting from the trigonometric Slavnov 
scalar product, we also derive the trigonometric version of 
$Z_{n \! \times \! n}$. 

Each determinant, $Z_{N \! \times \! N}$ and $Z_{n\! \times \! n}$, 
is a function of the set $\{x\}$ of cardinality $n$, associated with 
the $n$ horizontal lines, and the set $\{y\}$ of cardinality $N$, 
associated with the $N$ vertical lines.  
In Section {\bf \ref{section-4}}, as an independent check 
of the correctness of our expressions for $Z_{N\! \times \! N}$ 
and $Z_{n\! \times \! n}$, we show that they can be written as 
polynomials in each of their variables $x_i$, with the same bound
on their degree, and that they satisfy the same recursion relations 
and initial condition. This proves that they are equal, as expected 
from the fact that they are different expressions for 
the same partition function. 
In Section {\bf \ref{section-5}}, we recall basic facts regarding 
Casorati determinants (the discrete analogues of Wronskians)
and discrete KP $\tau$-functions, then we show that pDWPF's are 
discrete KP $\tau$-functions in the $\{x\}$ as well as in the 
$\{y\}$ variables.  

In Section {\bf \ref{section-6}}, we recall a mapping that Gromov 
and Vieira use in \cite{gromov.vieira.short, gromov.vieira.long} 
to introduce 1-loop corrections into 
the 0-loop expressions of certain structure constants in 
$\cN \! = \! 4$ supersymmetric Yang-Mills theory, and show that 
the $(n\! \times \! N)$ pDWPF remains a determinant under 
this mapping.
In Section {\bf \ref{section-7}}, we include remarks on recent 
developments.

\subsection{Glossary of frequently used notation}
$\{x\}$ $(\{y\})$ is a set of rapidity variables that do not 
satisfy Bethe equations and that flow along horizontal (vertical) 
lines. We always take $\{x\}$ and $\{y\}$ to be free variables.
$\{b\}$ is a set of rapidity variables that {\it do} satisfy the 
Bethe equations and that flow along horizontal lines. 
When a set $\{x\}$ has cardinality $N$, we sometimes indicate 
this by writing $\{x\}_N$. At times we also use the notation 

\begin{equation}
\Delta\{x\}_N 
=
\prod_{1 \leq i < j \leq N} [x_j-x_i],
\quad
\Delta\{-x\}_N 
=
\prod_{1 \leq i < j \leq N} [x_i-x_j]
\end{equation}

\noindent for Vandermonde determinants in the variables 
$\{x\}_N$. 
$[x - y] =       x - y $ in the rational      case, and  
$[x - y] = \sinh(x - y)$ in the trigonometric case.

%%%SEC01%%%
\section{Six-vertex model configurations}
\label{section-1}

In this section we recall basic definitions related to the six-vertex 
model on an $(n\! \times \! N)$ square lattice, $n \leq N$, including 
vertex model descriptions of Korepin's domain wall configurations on 
an $(N\! \times \! N)$ lattice \cite{korepin.paper}, Slavnov's scalar 
product configurations on a $(2n\! \times \! N)$ lattice \cite{wheeler}, 
and the determinant expressions for these objects 
\cite{izergin.paper, slavnov}. Finally, we define the partial domain 
wall configurations and the corresponding partial domain wall partition 
functions. 

\subsection{Lines, orientations and rapidity variables}
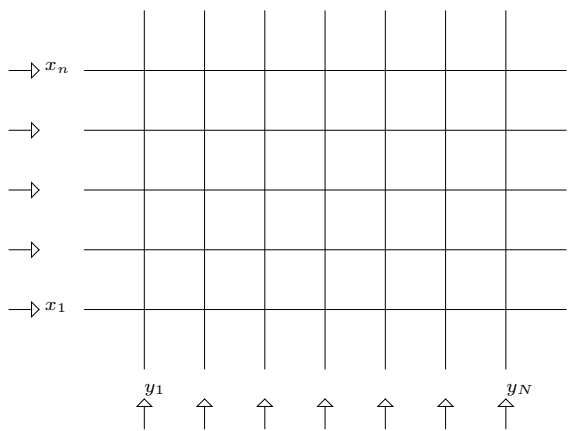
\begin{figure}[h]

\begin{center}
\begin{minipage}{4.3in}

\setlength{\unitlength}{0.0004cm}
\begin{picture}(20000,15500)(-8000,-5000)

\path(-2000,0)(14000,0)
\put(-3300,0){\scriptsize$x_{1}$}
\path(-4500,0)(-3500,0)
\whiten\path(-3750,250)(-3750,-250)(-3500,0)(-3750,250)

\path(-2000,2000)(14000,2000)
\path(-4500,2000)(-3500,2000)
\whiten\path(-3750,2250)(-3750,1750)(-3500,2000)(-3750,2250)

\path(-2000,4000)(14000,4000)
\path(-4500,4000)(-3500,4000)
\whiten\path(-3750,4250)(-3750,3750)(-3500,4000)(-3750,4250)

\path(-2000,6000)(14000,6000)
\path(-4500,6000)(-3500,6000)
\whiten\path(-3750,6250)(-3750,5750)(-3500,6000)(-3750,6250)

\path(-2000,8000)(14000,8000)
\put(-3300,8000){\scriptsize$x_n$}
\path(-4500,8000)(-3500,8000)
\whiten\path(-3750,8250)(-3750,7750)(-3500,8000)(-3750,8250)

%%%%%%%%%%%

\path(0,-2000)(0,10000)
\put(0,-2750){\scriptsize$y_1$}
\path(0,-4000)(0,-3000)
\whiten\path(-250,-3250)(250,-3250)(0,-3000)(-250,-3250)

\path(2000,-2000)(2000,10000)
\path(2000,-4000)(2000,-3000)
\whiten\path(1750,-3250)(2250,-3250)(2000,-3000)(1750,-3250)

\path(4000,-2000)(4000,10000)
\path(4000,-4000)(4000,-3000)
\whiten\path(3750,-3250)(4250,-3250)(4000,-3000)(3750,-3250)

\path(6000,-2000)(6000,10000)
\path(6000,-4000)(6000,-3000)
\whiten\path(5750,-3250)(6250,-3250)(6000,-3000)(5750,-3250)

\path(8000,-2000)(8000,10000)
\path(8000,-4000)(8000,-3000)
\whiten\path(7750,-3250)(8250,-3250)(8000,-3000)(7750,-3250)

\path(10000,-2000)(10000,10000)
\path(10000,-4000)(10000,-3000)
\whiten\path(9750,-3250)(10250,-3250)(10000,-3000)(9750,-3250)

\path(12000,-2000)(12000,10000)
\put(12000,-2750){\scriptsize$y_{N}$}
\path(12000,-4000)(12000,-3000)
\whiten\path(11750,-3250)(12250,-3250)(12000,-3000)(11750,-3250)

\end{picture}

\end{minipage}
\end{center}

\caption{A square $(n \! \times \! N)$ lattice with oriented lines 
and rapidity variables. The lines are assigned orientations indicated 
by the white arrows.}

\label{lattice}
\end{figure}

Consider a square lattice with $n$ horizontal lines and $N$ vertical 
lines that intersect at $(n \! \times \! N)$ points, $n \leq N$. We 
order the horizontal lines from bottom to top and assign the $i$-th 
line an orientation from left to right and a rapidity variable $x_i$. 
We order the vertical lines from left to right and assign
the $j$-th line an orientation from bottom to top and a rapidity 
variable $y_j$. See Figure {\bf \ref{lattice}}. 
The orientations that we assign to the lattice lines are matters 
of convention and are meant to make the vertices of the 
six-vertex model, that we introduce shortly, unambiguous. 

%FIG01

\subsection{Segments, arrows and vertices} 
Each lattice line is divided into segments by all other lines that 
are perpendicular to it. {\it Bulk segments}\/ are attached to two 
intersection points. {\it Boundary segments}\/ are attached to 
one intersection point only.
Assign each segment an arrow that can point in either direction, 
and define the vertex $v_{ij}$ as the union of 
the intersection point of the $i$-th horizontal line and the 
$j$-th vertical line, the four line segments attached to this 
intersection point, and the arrows on these segments. 

\subsection{Weights, configurations and partition functions} 
Assign every vertex $v_{ij}$ a weight $w_{ij}$ that depends on the 
specific orientations of its arrows, and the rapidities $x_i$ and 
$y_j$ that flow through it.
Any lattice configuration with a definite assignment of arrows is 
assigned a weight equal to the product of the weights of its vertices. 
The partition function of the lattice in Figure {\bf \ref{lattice}} 
is the sum of the weights of all lattice configurations which respect 
the boundary conditions that we impose. 

\subsection{Six vertices that conserve arrow flow}
\begin{figure}[h]
\begin{center}
\begin{minipage}{4.3in}
\setlength{\unitlength}{0.00038cm}
%(xwidth,ywidth)(-xpos,-ypos)
\begin{picture}(20000,16000)(-4500,-14000)
%
%%%Vertex a_{+}%%%
\path(-2000,0000)(2000,0000)
\blacken\path(-1250,250)(-1250,-250)(-750,0)(-1250,250)
\blacken\path(750,250)(750,-250)(1250,0)(750,250)
%%%
\put(-2750,0){\scriptsize{$x$}}
\path(-4000,0)(-3000,0)
\whiten\path(-3500,250)(-3500,-250)(-3000,0)(-3500,250)
%%%
\path(0000,-2000)(0000,2000)
\blacken\path(-250,-1250)(250,-1250)(0,-750)(-250,-1250)
\blacken\path(-250,750)(250,750)(0,1250)(-250,750)
%%%
\put(-1500,-5000){\scriptsize$a_{+}(x,y)$}
\put(0,-2750){\scriptsize{$y$}}
\path(0,-4000)(0,-3000)
\whiten\path(-250,-3500)(250,-3500)(0,-3000)(-250,-3500)
%
%%%Vertex b_{+}%%%
\path(8000,0000)(12000,0000)
\blacken\path(9250,250)(9250,-250)(8750,0)(9250,250)
\blacken\path(11250,250)(11250,-250)(10750,0)(11250,250)
%%%
\put(7250,0){\scriptsize{$x$}}
\path(6000,0)(7000,0)
\whiten\path(6500,250)(6500,-250)(7000,0)(6500,250)
%%%
\path(10000,-2000)(10000,2000)
\blacken\path(9750,-1250)(10250,-1250)(10000,-750)(9750,-1250)
\blacken\path(9750,750)(10250,750)(10000,1250)(9750,750)
%%%
\put(8500,-5000){\scriptsize$b_{+}(x,y)$}
\put(10000,-2750){\scriptsize{$y$}}
\path(10000,-4000)(10000,-3000)
\whiten\path(9750,-3500)(10250,-3500)(10000,-3000)(9750,-3500)
%
%%%Vertex c_{+}%%%
\path(18000,0000)(22000,0000)
\blacken\path(18750,250)(18750,-250)(19250,0)(18750,250)
\blacken\path(21250,250)(21250,-250)(20750,0)(21250,250)
%%%
\put(17250,0){\scriptsize{$x$}}
\path(16000,0)(17000,0)
\whiten\path(16500,250)(16500,-250)(17000,0)(16500,250)
%%%
\path(20000,-2000)(20000,2000)
\blacken\path(19750,-750)(20250,-750)(20000,-1250)(19750,-750)
\blacken\path(19750,750)(20250,750)(20000,1250)(19750,750)
%%%
\put(18500,-5000){\scriptsize$c_{+}(x,y)$}
\put(20000,-2750){\scriptsize{$y$}}
\path(20000,-4000)(20000,-3000)
\whiten\path(19750,-3500)(20250,-3500)(20000,-3000)(19750,-3500)
%
%%%Vertex a_{-}%%%
\path(-2000,-8000)(2000,-8000)
\blacken\path(-750,-7750)(-750,-8250)(-1250,-8000)(-750,-7750)
\blacken\path(1250,-7750)(1250,-8250)(750,-8000)(1250,-7750)
%%%
\put(-2750,-8000){\scriptsize{$x$}}
\path(-4000,-8000)(-3000,-8000)
\whiten\path(-3500,-7750)(-3500,-8250)(-3000,-8000)(-3500,-7750)
%%%
\path(0000,-10000)(0000,-6000)
\blacken\path(-250,-8750)(250,-8750)(0,-9250)(-250,-8750)
\blacken\path(-250,-6750)(250,-6750)(0,-7250)(-250,-6750)
%%%
\put(-1500,-13000){\scriptsize$a_{-}(x,y)$}
\put(0,-10750){\scriptsize{$y$}}
\path(0,-12000)(0,-11000)
\whiten\path(-250,-11500)(250,-11500)(0,-11000)(-250,-11500)
%
%%%Vertex b_{-}%%%
\path(8000,-8000)(12000,-8000)
\blacken\path(8750,-7750)(8750,-8250)(9250,-8000)(8750,-7750)
\blacken\path(10750,-7750)(10750,-8250)(11250,-8000)(10750,-7750)
%%%
\put(7250,-8000){\scriptsize{$x$}}
\path(6000,-8000)(7000,-8000)
\whiten\path(6500,-7750)(6500,-8250)(7000,-8000)(6500,-7750)
%%%
\path(10000,-10000)(10000,-6000)
\blacken\path(9750,-8750)(10250,-8750)(10000,-9250)(9750,-8750)
\blacken\path(9750,-6750)(10250,-6750)(10000,-7250)(9750,-6750)
%%%
\put(8500,-13000){\scriptsize$b_{-}(x,y)$}
\put(10000,-10750){\scriptsize{$y$}}
\path(10000,-12000)(10000,-11000)
\whiten\path(9750,-11500)(10250,-11500)(10000,-11000)(9750,-11500)
%
%%%Vertex c_{-}%%%
\path(18000,-8000)(22000,-8000)
\blacken\path(19250,-7750)(19250,-8250)(18750,-8000)(19250,-7750)
\blacken\path(20750,-7750)(20750,-8250)(21250,-8000)(20750,-7750)
%%%
\put(17250,-8000){\scriptsize{$x$}}
\path(16000,-8000)(17000,-8000)
\whiten\path(16500,-7750)(16500,-8250)(17000,-8000)(16500,-7750)
%%%
\path(20000,-10000)(20000,-6000)
\blacken\path(19750,-9250)(20250,-9250)(20000,-8750)(19750,-9250)
\blacken\path(19750,-6750)(20250,-6750)(20000,-7250)(19750,-6750)
%%%
\put(18500,-13000){\scriptsize$c_{-}(x,y)$}
\put(20000,-10750){\scriptsize{$y$}}
\path(20000,-12000)(20000,-11000)
\whiten\path(19750,-11500)(20250,-11500)(20000,-11000)(19750,-11500)
\end{picture}
\end{minipage}
\end{center}

\caption{Assignment of weights to vertices.} 

\label{six-vertices}
\end{figure}
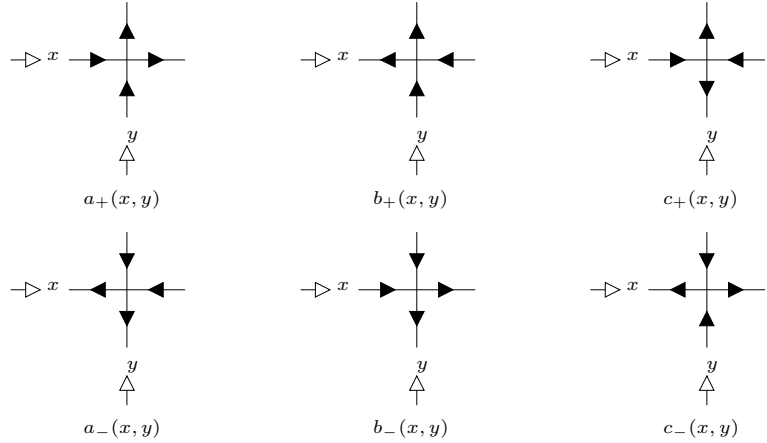

Since every arrow can point in either direction, there are $2^4 = 16$ 
possible (types of) vertices. We are interested in models with 
{\it \lq conservation of arrow flow\rq}. That is, the only vertices 
with non-zero weights are those such that the number of arrows that 
point toward the intersection point of the vertex is equal to the number 
of arrows that point away from it. These are six such vertices shown in 
Figure {\bf \ref{six-vertices}}. The remaining vertices have zero weights.

%FIG02 

In this work, we study the rational and the trigonometric 
six-vertex model. The former is a special case of the latter. 
For the rational six-vertex model, we use the weights

\begin{equation}
a_{\pm}(x,y) = 1,
\quad
b_{\pm}(x,y) = \frac{x-y}{x-y+1},
\quad
c_{\pm}(x,y) = \frac{1}{x-y+1}
\label{rat-wt}
\end{equation}

\noindent For the trigonometric six-vertex model, we use 
the weights

\begin{equation}
a_{\pm}(x,y) = 1,
\quad
b_{\pm}(x,y) = e^{\pm \gamma} \frac{[x-y]}{[x-y+\gamma]},
\quad
c_{\pm}(x,y) = e^{\pm (x-y)} \frac{[\gamma]}{[x-y+\gamma]}
\label{trig-wt}
\end{equation}

\noindent where $[x] \equiv \sinh (x)$. The weights in Equations 
({\bf \ref{rat-wt}}) and ({\bf \ref{trig-wt}}) satisfy the Yang-Baxter 
equations and unitarity. The parametrization in the trigonometric case 
is not unique. The reason for using the parametrization in Equation 
({\bf \ref{trig-wt}}) is explained in Subsection {\bf\ref{lim-DWPF-trig}}. 

\subsection{Limiting form of the weights}
From Equations ({\bf  \ref{rat-wt}}) and ({\bf \ref{trig-wt}}), 
as $x \rightarrow \infty$, the rational weights become

\begin{align}
a_{\pm}(x,y) \to 1,
\quad\quad
b_{\pm}(x,y) \to 1,
\quad\quad
c_{\pm}(x,y) \to \frac{1}{x}
\label{rat-lim}
\end{align}
while the trigonometric weights become
\begin{multline}
a_{\pm}(x,y) \sim b_{+}(x,y) \to 1,
\
b_{-}(x,y) \to e^{-2\gamma},
\ 
c_{+}(x,y) \to e^{-\gamma} [\gamma],
\
c_{-}(x,y) \to \frac{e^{-\gamma} [\gamma]}{e^{2(x-y)}}
\label{trig-lim}
\end{multline}

\subsection{The domain wall partition function, DWPF}

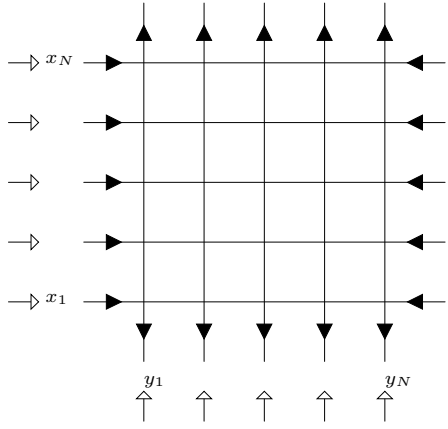
\begin{figure}[h]

\begin{center}
\begin{minipage}{4.3in}

\setlength{\unitlength}{0.0004cm}
\begin{picture}(20000,13500)(-10000,-4000)

\path(-2000,0)(10000,0)
\put(-3250,0){\scriptsize$x_1$}
\path(-4500,0)(-3500,0)
\whiten\path(-3750,250)(-3750,-250)(-3500,0)(-3750,250)
\blacken\path(-1250,250)(-1250,-250)(-750,0)(-1250,250)
\blacken\path(9250,250)(9250,-250)(8750,0)(9250,250)

\path(-2000,2000)(10000,2000)
\path(-4500,2000)(-3500,2000)
\whiten\path(-3750,2250)(-3750,1750)(-3500,2000)(-3750,2250)
\blacken\path(-1250,2250)(-1250,1750)(-750,2000)(-1250,2250)
\blacken\path(9250,2250)(9250,1750)(8750,2000)(9250,2250)

\path(-2000,4000)(10000,4000)
\path(-4500,4000)(-3500,4000)
\whiten\path(-3750,4250)(-3750,3750)(-3500,4000)(-3750,4250)
\blacken\path(-1250,4250)(-1250,3750)(-750,4000)(-1250,4250)
\blacken\path(9250,4250)(9250,3750)(8750,4000)(9250,4250)

\path(-2000,6000)(10000,6000)
\path(-4500,6000)(-3500,6000)
\whiten\path(-3750,6250)(-3750,5750)(-3500,6000)(-3750,6250)
\blacken\path(-1250,6250)(-1250,5750)(-750,6000)(-1250,6250)
\blacken\path(9250,6250)(9250,5750)(8750,6000)(9250,6250)

\path(-2000,8000)(10000,8000)
\put(-3250,8000){\scriptsize$x_N$}
\path(-4500,8000)(-3500,8000)
\whiten\path(-3750,8250)(-3750,7750)(-3500,8000)(-3750,8250)
\blacken\path(-1250,8250)(-1250,7750)(-750,8000)(-1250,8250)
\blacken\path(9250,8250)(9250,7750)(8750,8000)(9250,8250)

%%%%%%%%%%%

\path(0,-2000)(0,10000)
\put(0,-2750){\scriptsize$y_1$}
\path(0,-4000)(0,-3000)
\whiten\path(-250,-3250)(250,-3250)(0,-3000)(-250,-3250)
\blacken\path(-250,-750)(250,-750)(0,-1250)(-250,-750)
\blacken\path(-250,8750)(250,8750)(0,9250)(-250,8750)

\path(2000,-2000)(2000,10000)
\path(2000,-4000)(2000,-3000)
\whiten\path(1750,-3250)(2250,-3250)(2000,-3000)(1750,-3250)
\blacken\path(1750,-750)(2250,-750)(2000,-1250)(1750,-750)
\blacken\path(1750,8750)(2250,8750)(2000,9250)(1750,8750)

\path(4000,-2000)(4000,10000)
\path(4000,-4000)(4000,-3000)
\whiten\path(3750,-3250)(4250,-3250)(4000,-3000)(3750,-3250)
\blacken\path(3750,-750)(4250,-750)(4000,-1250)(3750,-750)
\blacken\path(3750,8750)(4250,8750)(4000,9250)(3750,8750)

\path(6000,-2000)(6000,10000)
\path(6000,-4000)(6000,-3000)
\whiten\path(5750,-3250)(6250,-3250)(6000,-3000)(5750,-3250)
\blacken\path(5750,-750)(6250,-750)(6000,-1250)(5750,-750)
\blacken\path(5750,8750)(6250,8750)(6000,9250)(5750,8750)

\path(8000,-2000)(8000,10000)
\put(8000,-2750){\scriptsize$y_N$}
\path(8000,-4000)(8000,-3000)
\whiten\path(7750,-3250)(8250,-3250)(8000,-3000)(7750,-3250)
\blacken\path(7750,-750)(8250,-750)(8000,-1250)(7750,-750)
\blacken\path(7750,8750)(8250,8750)(8000,9250)(7750,8750)

\end{picture}

\end{minipage}
\end{center}

\caption{Lattice definition of $Z(\{x\}_N | \{y\}_N)$. The boundary 
segments have the definite arrow assignments shown, and all bulk 
segments are summed over.}

\label{dwpf}
\end{figure}

This standard object is defined in six-vertex model terms 
as the partition function of the configurations in Figure 
{\bf \ref{dwpf}}, \cite{korepin.paper, korepin.book}. It 
depends on two sets of variables 
$\{x\}_N = \{x_1,    \dots,x_N\}$  and 
$\{y\}_N = \{y_1,    \dots,y_N\}$, and we denote it by 
$Z( \{x\}_N | \{y\}_N )$. 

%FIG03

\subsection{Izergin's determinant}

Following \cite{izergin.paper}, in the rational parametrization of 
Equation ({\bf \ref{rat-wt}}) the DWPF is given by 

\begin{align}
Z \ll \{x\}_N \Big| \{y\}_N \rr
=
\frac{\displaystyle{
\prod_{i,j=1}^{N} (x_i - y_j)
}}
{\displaystyle{
\Delta\{x\}_N \Delta\{-y\}_N
%\prod_{1 \leq i < j \leq N} (x_j - x_i)(y_i - y_j)
}}
\det
\ll
\frac{1}{(x_i - y_j)(x_i - y_j + 1)}
\rr_{1\leq i,j \leq N}
\label{IK}
\end{align}
In the trigonometric parametrization of 
Equation ({\bf \ref{trig-wt}}), the DWPF is given by

\begin{align}
Z \ll \{x\}_N \Big| \{y\}_N \rr
=
\frac{\displaystyle{
e^{|x|-|y|}
\prod_{i,j=1}^{N} [x_i - y_j]
}}
{\displaystyle{
\Delta\{x\}_N \Delta\{-y\}_N
%\prod_{1 \leq i < j \leq N} [x_j - x_i][y_i - y_j]
}}
\det
\ll
\frac{[\gamma]}{[x_i - y_j][x_i - y_j + \gamma]}
\rr_{1\leq i,j \leq N}
\label{IK-trig}
\end{align}

\noindent where we use the notation $|x|= \sum_{k=1}^{N} x_k$.

\subsection{The scalar product}

\begin{figure}[h]

\begin{center}
\begin{minipage}{4.3in}

\setlength{\unitlength}{0.0004cm}
\begin{picture}(20000,15500)(-10000,-8000)

%%%%%%%

\path(-2000,-4000)(10000,-4000)
\put(-3250,-4000){\scriptsize$x_1$}
\path(-4500,-4000)(-3500,-4000)
\whiten\path(-3750,-3750)(-3750,-4250)(-3500,-4000)(-3750,-3750)
\blacken\path(-750,-3750)(-750,-4250)(-1250,-4000)(-750,-3750)
\blacken\path(8750,-3750)(8750,-4250)(9250,-4000)(8750,-3750)

\path(-2000,-2000)(10000,-2000)
\path(-4500,-2000)(-3500,-2000)
\whiten\path(-3750,-1750)(-3750,-2250)(-3500,-2000)(-3750,-1750)
\blacken\path(-750,-1750)(-750,-2250)(-1250,-2000)(-750,-1750)
\blacken\path(8750,-1750)(8750,-2250)(9250,-2000)(8750,-1750)

\path(-2000,0)(10000,0)
\put(-3250,0){\scriptsize$x_n$}
\path(-4500,0)(-3500,0)
\whiten\path(-3750,250)(-3750,-250)(-3500,0)(-3750,250)
\blacken\path(-750,250)(-750,-250)(-1250,0)(-750,250)
\blacken\path(8750,250)(8750,-250)(9250,0)(8750,250)

%%%%%%%%

%%%%%%%%

\path(-2000,2000)(10000,2000)
\put(-3250,2000){\scriptsize$b_1$}
\path(-4500,2000)(-3500,2000)
\whiten\path(-3750,2250)(-3750,1750)(-3500,2000)(-3750,2250)
\blacken\path(-1250,2250)(-1250,1750)(-750,2000)(-1250,2250)
\blacken\path(9250,2250)(9250,1750)(8750,2000)(9250,2250)

\path(-2000,4000)(10000,4000)
\path(-4500,4000)(-3500,4000)
\whiten\path(-3750,4250)(-3750,3750)(-3500,4000)(-3750,4250)
\blacken\path(-1250,4250)(-1250,3750)(-750,4000)(-1250,4250)
\blacken\path(9250,4250)(9250,3750)(8750,4000)(9250,4250)

\path(-2000,6000)(10000,6000)
\put(-3250,6000){\scriptsize$b_n$}
\path(-4500,6000)(-3500,6000)
\whiten\path(-3750,6250)(-3750,5750)(-3500,6000)(-3750,6250)
\blacken\path(-1250,6250)(-1250,5750)(-750,6000)(-1250,6250)
\blacken\path(9250,6250)(9250,5750)(8750,6000)(9250,6250)

%%%%%%%%%%%

%%%%%%%%%%%

\path(0,-6000)(0,8000)
\put(-250,-6750){\scriptsize$y_1$}
\path(0,-8250)(0,-7250)
\whiten\path(-250,-7500)(250,-7500)(0,-7250)(-250,-7500)
\blacken\path(-250,-5250)(250,-5250)(0,-4750)(-250,-5250)
\blacken\path(-250,6750)(250,6750)(0,7250)(-250,6750)

\path(2000,-6000)(2000,8000)
\path(2000,-8250)(2000,-7250)
\whiten\path(1750,-7500)(2250,-7500)(2000,-7250)(1750,-7500)
\blacken\path(1750,-5250)(2250,-5250)(2000,-4750)(1750,-5250)
\blacken\path(1750,6750)(2250,6750)(2000,7250)(1750,6750)

\path(4000,-6000)(4000,8000)
\path(4000,-8250)(4000,-7250)
\whiten\path(3750,-7500)(4250,-7500)(4000,-7250)(3750,-7500)
\blacken\path(3750,-5250)(4250,-5250)(4000,-4750)(3750,-5250)
\blacken\path(3750,6750)(4250,6750)(4000,7250)(3750,6750)

\path(6000,-6000)(6000,8000)
\path(6000,-8250)(6000,-7250)
\whiten\path(5750,-7500)(6250,-7500)(6000,-7250)(5750,-7500)
\blacken\path(5750,-5250)(6250,-5250)(6000,-4750)(5750,-5250)
\blacken\path(5750,6750)(6250,6750)(6000,7250)(5750,6750)

\path(8000,-6000)(8000,8000)
\put(7750,-6750){\scriptsize$y_N$}
\path(8000,-8250)(8000,-7250)
\whiten\path(7750,-7500)(8250,-7500)(8000,-7250)(7750,-7500)
\blacken\path(7750,-5250)(8250,-5250)(8000,-4750)(7750,-5250)
\blacken\path(7750,6750)(8250,6750)(8000,7250)(7750,6750)

\end{picture}

\end{minipage}
\end{center}

\caption{Lattice representation of 
$S ( \{x\}_n, \{b\}_n | \{y\}_N )$. 
There are two sets of horizontal rapidities $\{x\}_n$ and $\{b\}_n$, 
and one set of vertical rapidities $\{y\}_N$. The variables $\{b\}_n$ 
satisfy Bethe equations.}
\label{scalar-prod}
\end{figure}
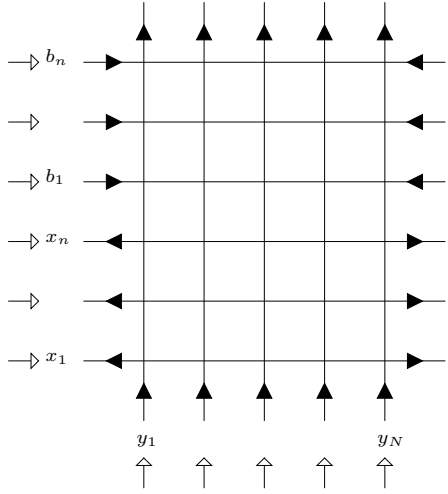

This is another standard object that is defined in this work in 
six-vertex model terms
\footnote{\  The scalar product is usually 
defined as a vacuum expectation value of algebraic Bethe Ansatz 
operators. For our purposes, this formalism in unnecessary.}
as the partition function of the configuration in 
Figure {\bf \ref{scalar-prod}}, \cite{slavnov, KMT, wheeler}. 
It depends on three sets of variables 
$\{x\}_n = \{x_1,\dots,x_n\}$, 
$\{b\}_n = \{b_1,\dots,b_n\}$, 
$\{y\}_N = \{y_1,\dots,y_N\}$, 
where $n \leq N$. We denote it by $S ( \{x\}_n,\{b\}_n | \{y\}_N )$.

%FIG04

\subsection{Slavnov's determinant}
Following \cite{slavnov} we assume that one set of variables in 
Figure {\bf \ref{scalar-prod}}, $\{b\}_n$, obeys the Bethe equations
\footnote{\  In this work, all rapidity variables denoted 
by $b_i$ are assumed to obey Bethe equations, while all rapidity 
variables denoted by $x_i$ or $y_j$ are free.}. In the rational 
and trigonometric parametrizations they are given by

\begin{multline}
\prod_{j=1}^{N}
\ll
\frac{b_i-y_j+1}{b_i-y_j}
\rr
=
\prod_{j \not= i}^{n}
\ll
\frac{b_i-b_j + 1}{b_i - b_j - 1},
\rr
\\
\prod_{j=1}^{N}\frac{[b_i-y_j+\gamma]}{[b_i-y_j]}
=
e^{N \gamma}
\prod_{j \not= i}^{n}
\frac{[b_i-b_j + \gamma]}{[b_i - b_j - \gamma]},
\ \ 
\forall\ 1 \leq i \leq n
\label{bethe}
\end{multline}
respectively. Assuming that the Bethe equations hold, in the rational 
parametrization the scalar product has the determinant representation

\begin{multline}
\label{betscal}
S\ll \{x\}_n, \{b\}_n \Big| \{y\}_N\rr
=
\Delta^{-1}\{x\}_n \Delta^{-1}\{-b\}_n
%\frac{\displaystyle{
%1
%}}
%{\displaystyle{
%\prod_{1\leq i < j \leq n}
%(x_j-x_i)(b_i-b_j)
%}}
\\
\times
\det 
\ll
\frac{\displaystyle{
\prod_{k\not= j}^{n}(b_k-x_i-1)
\prod_{k=1}^{N}
\ll
\frac{x_i-y_k}{x_i-y_k+1}
\rr
-
\prod_{k\not=j}^{n} (b_k-x_i+1)
}}
{x_i-b_j}
\rr_{1\leq i,j \leq n}
\end{multline}

\noindent In the trigonometric parametrization, it is given by

\begin{multline}
S\ll \{x\}_n, \{b\}_n \Big| \{y\}_N\rr
=
%\frac{\displaystyle{
[\gamma]^n e^{|b|-|x|}
\Delta^{-1}\{x\}_n \Delta^{-1}\{-b\}_n
%}}
%{\displaystyle{
%\prod_{1\leq i < j \leq n}
%[x_j-x_i][b_i-b_j]
%}}
\label{betscal2}
\\
\times
\det 
\ll
\frac{\displaystyle{
e^{(N-n)\gamma}
\prod_{k\not= j}^{n}[b_k-x_i-\gamma]
\prod_{k=1}^{N}
\frac{[x_i-y_k]}{[x_i-y_k+\gamma]}
-
e^{-n\gamma}
\prod_{k\not=j}^{n} [b_k-x_i+\gamma]
}}
{[x_i-b_j]}
\rr_{1\leq i,j \leq n}
\end{multline}

\subsection{The partial domain wall partition function, pDWPF} 
\begin{figure}[h]

\begin{center}
\begin{minipage}{4.3in}

\setlength{\unitlength}{0.0004cm}
\begin{picture}(20000,9500)(-21000,-3400)

\path(-2000,0)(10000,0)
\put(-3250,0){\scriptsize$x_1$}
\path(-4500,0)(-3500,0)
\whiten\path(-3750,250)(-3750,-250)(-3500,0)(-3750,250)
\blacken\path(-750,250)(-750,-250)(-1250,0)(-750,250)
\blacken\path(8750,250)(8750,-250)(9250,0)(8750,250)

\path(-2000,2000)(10000,2000)
\path(-4500,2000)(-3500,2000)
\whiten\path(-3750,2250)(-3750,1750)(-3500,2000)(-3750,2250)
\blacken\path(-750,2250)(-750,1750)(-1250,2000)(-750,2250)
\blacken\path(8750,2250)(8750,1750)(9250,2000)(8750,2250)

\path(-2000,4000)(10000,4000)
\put(-3250,4000){\scriptsize$x_n$}
\path(-4500,4000)(-3500,4000)
\whiten\path(-3750,4250)(-3750,3750)(-3500,4000)(-3750,4250)
\blacken\path(-750,4250)(-750,3750)(-1250,4000)(-750,4250)
\blacken\path(8750,4250)(8750,3750)(9250,4000)(8750,4250)

%%%%%%%%%%%

\path(0,-2000)(0,6000)
\put(0,-2750){\scriptsize$y_1$}
\path(0,-4000)(0,-3000)
\whiten\path(-250,-3250)(250,-3250)(0,-3000)(-250,-3250)
\blacken\path(-250,-1250)(250,-1250)(0,-750)(-250,-1250)

\path(2000,-2000)(2000,6000)
\path(2000,-4000)(2000,-3000)
\whiten\path(1750,-3250)(2250,-3250)(2000,-3000)(1750,-3250)
\blacken\path(1750,-1250)(2250,-1250)(2000,-750)(1750,-1250)

\path(4000,-2000)(4000,6000)
\path(4000,-4000)(4000,-3000)
\whiten\path(3750,-3250)(4250,-3250)(4000,-3000)(3750,-3250)
\blacken\path(3750,-1250)(4250,-1250)(4000,-750)(3750,-1250)

\path(6000,-2000)(6000,6000)
\path(6000,-4000)(6000,-3000)
\whiten\path(5750,-3250)(6250,-3250)(6000,-3000)(5750,-3250)
\blacken\path(5750,-1250)(6250,-1250)(6000,-750)(5750,-1250)

\path(8000,-2000)(8000,6000)
\put(8000,-2750){\scriptsize$y_N$}
\path(8000,-4000)(8000,-3000)
\whiten\path(7750,-3250)(8250,-3250)(8000,-3000)(7750,-3250)
\blacken\path(7750,-1250)(8250,-1250)(8000,-750)(7750,-1250)

%%%%%%%%%%%%%%%%%%

\path(-22000,0)(-10000,0)
\put(-23250,0){\scriptsize$x_1$}
\path(-24500,0)(-23500,0)
\whiten\path(-23750,250)(-23750,-250)(-23500,0)(-23750,250)
\blacken\path(-21250,250)(-21250,-250)(-20750,0)(-21250,250)
\blacken\path(-10750,250)(-10750,-250)(-11250,0)(-10750,250)

\path(-22000,2000)(-10000,2000)
\path(-24500,2000)(-23500,2000)
\whiten\path(-23750,2250)(-23750,1750)(-23500,2000)(-23750,2250)
\blacken\path(-21250,2250)(-21250,1750)(-20750,2000)(-21250,2250)
\blacken\path(-10750,2250)(-10750,1750)(-11250,2000)(-10750,2250)

\path(-22000,4000)(-10000,4000)
\put(-23250,4000){\scriptsize$x_n$}
\path(-24500,4000)(-23500,4000)
\whiten\path(-23750,4250)(-23750,3750)(-23500,4000)(-23750,4250)
\blacken\path(-21250,4250)(-21250,3750)(-20750,4000)(-21250,4250)
\blacken\path(-10750,4250)(-10750,3750)(-11250,4000)(-10750,4250)

%%%%%%%%%%%

\path(-20000,-2000)(-20000,6000)
\put(-20000,-2750){\scriptsize$y_1$}
\path(-20000,-4000)(-20000,-3000)
\whiten\path(-20250,-3250)(-19750,-3250)(-20000,-3000)(-20250,-3250)
\blacken\path(-20250,-750)(-19750,-750)(-20000,-1250)(-20250,-750)

\path(-18000,-2000)(-18000,6000)
\path(-18000,-4000)(-18000,-3000)
\whiten\path(-18250,-3250)(-17750,-3250)(-18000,-3000)(-18250,-3250)
\blacken\path(-18250,-750)(-17750,-750)(-18000,-1250)(-18250,-750)

\path(-16000,-2000)(-16000,6000)
\path(-16000,-4000)(-16000,-3000)
\whiten\path(-16250,-3250)(-15750,-3250)(-16000,-3000)(-16250,-3250)
\blacken\path(-16250,-750)(-15750,-750)(-16000,-1250)(-16250,-750)

\path(-14000,-2000)(-14000,6000)
\path(-14000,-4000)(-14000,-3000)
\whiten\path(-14250,-3250)(-13750,-3250)(-14000,-3000)(-14250,-3250)
\blacken\path(-14250,-750)(-13750,-750)(-14000,-1250)(-14250,-750)

\path(-12000,-2000)(-12000,6000)
\put(-12000,-2750){\scriptsize$y_N$}
\path(-12000,-4000)(-12000,-3000)
\whiten\path(-12250,-3250)(-11750,-3250)(-12000,-3000)(-12250,-3250)
\blacken\path(-12250,-750)(-11750,-750)(-12000,-1250)(-12250,-750)

\end{picture}

\end{minipage}
\end{center}

\caption{On the left, lattice representation of $Z_1( \{x\}_n | \{y\}_N )$. 
On the right, lattice representation of $Z_2( \{x\}_n | \{y\}_N )$. 
The number of horizontal rapidities $x_i$ is less than the number 
of vertical rapidities $y_j$. The top boundary segments are without 
arrows to indicate summation at these points.}

\label{dwpf-res}
\end{figure}
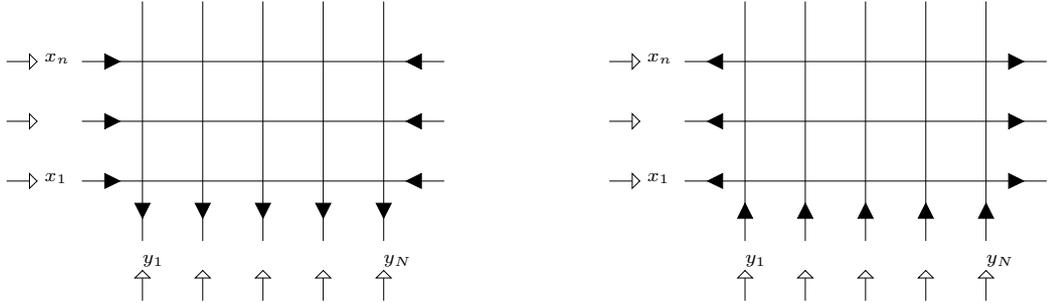

Let $n$ be an integer satisfying $1 \leq n \leq N $. Consider 
the partition function generated by deleting the top $(N-n)$ 
rows from the lattice in Figure {\bf \ref{dwpf}}, or the top $n$ 
rows from the lattice in Figure {\bf \ref{scalar-prod}}, and 
whose top boundary is summed over all arrow configurations. 
We denote these objects by $Z_1 ( \{x\}_n | \{y\}_N )$ and 
$Z_2 ( \{x\}_n | \{y\}_N )$ respectively, and represent
them by the lattices in Figure {\bf \ref{dwpf-res}}.

%FIG05

We emphasize that, unlike the usual domain wall configurations, the top 
boundary segments in Figure {\bf \ref{dwpf-res}} are {\it not}\/ fixed 
to definite arrow configurations but are summed over just as the bulk 
segments.

As we will see in Subsection {\bf \ref{lim-DWPF-rat}}, up to a numerical 
coefficient, $Z_1(\{x\}_n | \{y\}_N)$ is the leading term in 
$Z(\{x\}_N | \{y\}_N)$ as $x_N,\dots,x_{n+1} \rightarrow \infty$. 
In this limit, the contribution from the top $(N-n)$ rows of 
Figure {\bf\ref{dwpf}} 
becomes trivial, and we are left with the lattice shown in 
Figure {\bf\ref{dwpf-res}}. 
For this reason, $Z_1(\{x\}_n | \{y\}_N)$ is a {\it partial domain 
wall partition function}, pDWPF.

One can also calculate the pDWPF $Z_2( \{x\}_n | \{y\}_N )$ as 
the leading term of the scalar product 
$S ( \{x\}_n,\{b\}_n | \{y\}_N )$ as 
$b_n,\dots,b_1 \rightarrow \infty$. 
In this limit, the contribution from the top $n$ rows of 
Figure {\bf\ref{scalar-prod}} becomes trivial, and we are left 
with the lattice shown in Figure {\bf\ref{dwpf-res}}. This is 
discussed in Subsection {\bf\ref{lim-sp}}.

In this paper we calculate $Z_1( \{x\}_n | \{y\}_N)$ and 
$Z_2(\{x\}_n | \{y\}_N)$ by taking the two limits described above. 
The starting points for these calculations are, respectively, Izergin's 
determinant formula for 
$Z( \{x\}_N | \{y\}_N)$ and Slavnov's determinant formula for 
$S( \{x\}_n,\{b\}_n | \{y\}_N)$. In the case of the rational six-vertex 
model, whose vertex weights are invariant under the reversal of all arrows, 
the two quantities $Z_1( \{x\}_n | \{y\}_N)$ and $Z_2(\{x\}_n | \{y\}_N)$ 
are in fact equal (which is easily verified by comparing the two lattices 
in Figure {\bf \ref{dwpf-res}}). Therefore in the rational parametrization 
we obtain two different determinant expressions for the same object.   

\subsection{Deleting lines from opposite boundaries.} 
By symmetry of $Z ( \{x\}_N | \{y\}_N )$ in the variables $\{x\}_N$, we are 
free to distribute the rapidities $x_N, \dots, x_{n+1}$ over the horizontal 
lines of the lattice in any way we wish, prior to taking the limit 
$x_N, \dots, x_{n+1} \rightarrow \infty$.

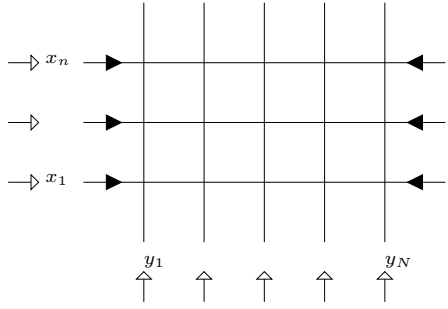
\begin{figure}[h]

\begin{center}
\begin{minipage}{4.3in}

\setlength{\unitlength}{0.0004cm}
\begin{picture}(20000,9500)(-10000,-4000)

\path(-2000,0)(10000,0)
\put(-3250,0){\scriptsize$x_1$}
\path(-4500,0)(-3500,0)
\whiten\path(-3750,250)(-3750,-250)(-3500,0)(-3750,250)
\blacken\path(-1250,250)(-1250,-250)(-750,0)(-1250,250)
\blacken\path(9250,250)(9250,-250)(8750,0)(9250,250)

\path(-2000,2000)(10000,2000)
\path(-4500,2000)(-3500,2000)
\whiten\path(-3750,2250)(-3750,1750)(-3500,2000)(-3750,2250)
\blacken\path(-1250,2250)(-1250,1750)(-750,2000)(-1250,2250)
\blacken\path(9250,2250)(9250,1750)(8750,2000)(9250,2250)

\path(-2000,4000)(10000,4000)
\put(-3250,4000){\scriptsize$x_n$}
\path(-4500,4000)(-3500,4000)
\whiten\path(-3750,4250)(-3750,3750)(-3500,4000)(-3750,4250)
\blacken\path(-1250,4250)(-1250,3750)(-750,4000)(-1250,4250)
\blacken\path(9250,4250)(9250,3750)(8750,4000)(9250,4250)

%%%%%%%%%%%

\path(0,-2000)(0,6000)
\put(0,-2750){\scriptsize$y_1$}
\path(0,-4000)(0,-3000)
\whiten\path(-250,-3250)(250,-3250)(0,-3000)(-250,-3250)
%\blacken\path(-250,4750)(250,4750)(0,5250)(-250,4750)

\path(2000,-2000)(2000,6000)
\path(2000,-4000)(2000,-3000)
\whiten\path(1750,-3250)(2250,-3250)(2000,-3000)(1750,-3250)
%\blacken\path(1750,4750)(2250,4750)(2000,5250)(1750,4750)

\path(4000,-2000)(4000,6000)
\path(4000,-4000)(4000,-3000)
\whiten\path(3750,-3250)(4250,-3250)(4000,-3000)(3750,-3250)
%\blacken\path(3750,4750)(4250,4750)(4000,5250)(3750,4750)

\path(6000,-2000)(6000,6000)
\path(6000,-4000)(6000,-3000)
\whiten\path(5750,-3250)(6250,-3250)(6000,-3000)(5750,-3250)
%\blacken\path(5750,4750)(6250,4750)(6000,5250)(5750,4750)

\path(8000,-2000)(8000,6000)
\put(8000,-2750){\scriptsize$y_N$}
\path(8000,-4000)(8000,-3000)
\whiten\path(7750,-3250)(8250,-3250)(8000,-3000)(7750,-3250)
%\blacken\path(7750,4750)(8250,4750)(8000,5250)(7750,4750)

\end{picture}

\end{minipage}
\end{center}

\caption{An alternative lattice representation of 
$Z_1 ( \{x\}_n | \{y\}_N )$.  Horizontal lines get removed from the top 
and bottom of the DWPF lattice. Both the top and bottom boundary segments 
are without arrows to indicate summation at these points. The top boundary 
is summed over all configurations which have exactly $(m-n)$ downward facing 
arrows, while the bottom boundary is summed over all configurations with 
exactly $(N-m)$ upward facing arrows. This lattice sum is equal to the one 
on the left of Figure {\bf\ref{dwpf-res}}, up to an overall factor.}

\label{dwpf-res-alt}
\end{figure}

For example, we can choose to place the variables $x_N,\dots,x_{m+1}$ on
the lowest lines of the lattice and $x_m,\dots,x_{n+1}$ on the highest,
where $m$ is some integer satisfying $n \leq m \leq N$. In the limit
$x_N,\dots,x_{n+1} \rightarrow \infty$, the bottom $(N-m)$ and the top
$(m-n)$ rows become trivial, and we obtain the lattice shown in
Figure {\bf\ref{dwpf-res-alt}}.

%FIG06
The lattice sum in Figure {\bf\ref{dwpf-res-alt}} is equal to the one on 
the left of Figure {\bf\ref{dwpf-res}}, up to multiplication by an overall 
factor. In the rational six-vertex model, this factor is the binomial 
coefficient $\binom{N-n}{N-m}$.

%%%SEC02%%%
\section{Domain wall partition function in the infinite-rapidity limit}
\label{section-2}

In this section, we obtain a determinant expression for the pDWPF 
starting from Izergin's formula for the DWPF, Equation ({\bf\ref{IK}}), 
and taking appropriate limits. The $(N\! \times \! N)$ determinant that 
we obtain is {\it\lq hybrid\rq} in the sense that it contains $n$ rows 
of the type in Izergin's formula, and $(N-n)$ rows of Vandermonde 
determinant-type. 

\subsection{One rapidity becomes infinite}

Consider the rational DWPF. Due to the domain wall boundary conditions, 
and the conservation of arrow flow, the top row of the lattice always 
contains precisely one $c_{+}$ vertex, while all remaining vertices in 
that row are of the type $a_{+}$ or $b_{+}$. 
Using the asymptotic behaviour of the vertex weights 
in Equation ({\bf \ref{rat-lim}}) and the definition of 
$Z_1( \{x\}_{N-1} | \{y\}_N)$, it is easy to see that

\begin{align}
Z_1\ll \{x\}_N \Big| \{y\}_N\rr
\to
\frac{Z_1 \ll \{x\}_{N-1} \Big| \{y\}_N \rr}{x_N}, 
\quad
\text{as}\ x_N \rightarrow \infty
\end{align}

\noindent and the pDWPF $Z_1 ( \{x\}_{N-1} | \{y\}_N )$ can be computed 
from the DWPF as

\begin{align}
Z_1 \ll \{x\}_{N-1} \Big| \{y\}_N\rr
=
\lim_{x_N \rightarrow \infty}
\ll
x_N
Z_1 \ll \{x\}_N \Big| \{y\}_N\rr
\rr
\label{one-limit}
\end{align}

\subsection{$(N-n)$ rapidities become infinite}
\label{lim-DWPF-rat}

Consider the lattice representation of the pDWPF $Z_1 ( \{x\}_i | \{y\}_N )$, 
for some $1 \leq i \leq N$. The top boundary of this lattice consists of 
a sum over $\binom{N}{i}$ possible arrow configurations. From now on, we 
make remarks which apply to the internal part of this lattice, assuming 
that the boundary is fixed to any one of these $\binom{N}{i}$ configurations.

Consider the $x_i$-row of vertices
\footnote{\  All vertices through 
which the $x_i$ rapidity variable flows.}. 
Any configuration that this row takes must contain at least one $c_{+}$ 
vertex and $m$ pairs of $\{c_+, c_-\}$ vertices, $m = 0, 1, 2, \cdots$
In the large $x_i$ limit, the leading contribution to 
$Z_1 ( \{x\}_i | \{y\}_N )$ corresponds to $m=0$. 
Taking multiple counting into consideration, we obtain 

\begin{align}
Z_1\ll \{x\}_i \Big| \{y\}_N\rr
\to
(N-i+1)
\frac{Z_1 \ll \{x\}_{i-1} \Big| \{y\}_N \rr}{x_i}, 
\quad
\text{as}\ 
x_i \rightarrow \infty
\end{align}
Iterating this result through $i=\{N,\dots,n+1\}$ we obtain 

\begin{align}
Z_1 \ll \{x\}_n \Big| \{y\}_N \rr
=
\frac{1}{(N-n)!}
\lim_{x_N,\dots,x_{n+1} \rightarrow \infty}
\ll
x_{n+1} \cdots x_{N}
Z_1\ll \{x\}_N \Big| \{y\}_N\rr
\rr
\label{many-limits}
\end{align}

\noindent where the limits are to be taken sequentially, starting 
with $x_N$.

\subsection{Limit of Izergin's determinant as one rapidity becomes 
infinite}

Starting from the expression in Equation ({\bf \ref{IK}}) for 
$Z_1 ( \{x\}_N | \{y\}_N )$, it is simple to take the limit 
specified in Equation ({\bf \ref{one-limit}}). Absorbing the 
factor 
$
x_N \prod_{j=1}^{N} (x_N - y_j)/
\prod_{j=1}^{N-1} (x_N - x_j)
$
into the final row of the determinant in Equation ({\bf \ref{IK}}), 
writing $x_N = 1/ \epsilon$ and taking $\epsilon \rightarrow 0$, we 
get 

\begin{multline}
\lim_{x_N \rightarrow \infty}
\ll
x_N
Z_1 \ll \{x\}_N \Big| \{y\}_N \rr
\rr
=
\prod_{i=1}^{N-1} \prod_{j=1}^{N} (x_i - y_j)
\Delta^{-1}\{x\}_{N-1}
\Delta^{-1}\{-y\}_N
\\
\times
\lim_{\epsilon \rightarrow 0}
\left|
\begin{array}{ccc}
\frac{1}{(x_1 - y_1)(x_1 - y_1 + 1)} 
& 
\cdots 
& 
\frac{1}{(x_1 - y_N)(x_1 - y_N + 1)}
\\
\vdots 
& & 
\vdots
\\
\frac{1}{(x_{N-1} - y_1)(x_{N-1} - y_1 + 1)}
&
\cdots
&
\frac{1}{(x_{N-1} - y_N)(x_{N-1} - y_N + 1)}
\\
\\
f^{(N)}_1(\epsilon)
&
\cdots
&
f^{(N)}_N(\epsilon)
\end{array}
\right|
\label{IK-1}
\end{multline}
where for all $1\leq i \leq N$ we have defined the function

\begin{align}
f^{(i)}_j(\epsilon)
=
\frac{\prod_{k=1}^{N} 
 (1-y_k \epsilon)}
{(1-y_j \epsilon) (1 - \bary_j \epsilon) 
\prod_{k=1}^{i-1} (1 -     x_k \epsilon)}
\label{aux-funct}
\end{align}
and set $\bary_j = y_j -1$ for convenience. In the limit, 
every entry of the final row goes to 1, hence

\begin{multline}
\lim_{x_N \rightarrow \infty}
\ll
x_N
Z_1 \ll \{x\}_N \Big| \{y\}_N \rr
\rr
=
\prod_{i=1}^{N-1} \prod_{j=1}^{N} (x_i - y_j)
\Delta^{-1}\{x\}_{N-1} \Delta^{-1}\{-y\}_N
\\
\times
\left|
\begin{array}{ccc}
\frac{1}{(x_1 - y_1)(x_1 - y_1 + 1)} 
& 
\cdots 
& 
\frac{1}{(x_1 - y_N)(x_1 - y_N + 1)}
\\
\vdots 
& & 
\vdots
\\
\frac{1}{(x_{N-1} - y_1)(x_{N-1} - y_1 + 1)}
&
\cdots
&
\frac{1}{(x_{N-1} - y_N)(x_{N-1} - y_N + 1)}
\\
\\
1
&
\cdots
&
1
\end{array}
\right|
\label{IK-2}
\end{multline}

\subsection{Limit of Izergin's determinant as $(N-n)$ rapidities become 
infinite}

\begin{myLemma}
If $h_i(y_j,\bar{y}_j)$ is the $i$-th complete symmetric function in 
two variables $y_j,\bar{y}_j$, given by the generating series

\begin{align}
\sum_{i=0}^{\infty}
h_i(y_j,\bar{y}_j)
\epsilon^i
=
\frac{1}{(1-y_j \epsilon)(1-\bar{y}_j \epsilon)}
\end{align}

\noindent then
\begin{multline}
\lim_{x_N,\dots,x_{n+1} \rightarrow \infty}
\ll
x_{n+1} \cdots x_{N}
Z_1 \ll \{x\}_N \Big| \{y\}_N \rr
\rr
=
\\
\prod_{i=1}^{n} 
\prod_{j=1}^{N}
(x_i - y_j)
\Delta^{-1}\{x\}_n \Delta^{-1}\{-y\}_N
\left|
\begin{array}{ccc}
\frac{1}{(x_1 - y_1)(x_1 - y_1 + 1)} 
& 
\cdots 
& 
\frac{1}{(x_1 - y_N)(x_1 - y_N + 1)}
\\
\vdots 
& & 
\vdots
\\
\frac{1}{(x_{n} - y_1)(x_{n} - y_1 + 1)}
&
\cdots
&
\frac{1}{(x_{n} - y_N)(x_{n} - y_N + 1)}
\\
\\
h_{N-n-1}(y_1,\bary_1) & \cdots & h_{N-n-1}(y_N,\bary_N)
\\
\vdots & & \vdots
\\
h_0(y_1,\bary_1) & \cdots & h_0(y_N,\bary_N)
\end{array}
\right|
\label{conjecture}
\end{multline}
\end{myLemma}
\myProof
Let $\mathcal{P}_{N-n}$ denote the proposition that 
Equation ({\bf\ref{conjecture}}) is true. Based on 
Equation ({\bf\ref{IK-2}}) for the one-rapidity case, 
we see that $\mathcal{P}_1$ is true. Let us assume that 
$\mathcal{P}_{N-n}$ is true and show that this implies 
$\mathcal{P}_{N-n+1}$. 
Multiplying Equation ({\bf\ref{conjecture}}) by $x_n$, 
making the change of variables $x_n = 1/ \epsilon$ and 
taking the limit $\epsilon \rightarrow 0$ gives

\begin{multline}
\lim_{x_N, \dots, x_n \rightarrow \infty}
\ll
x_n \cdots x_N
Z_1 \ll \{x\}_N \Big| \{y\}_N \rr
\rr
=
\prod_{i=1}^{n-1} 
\prod_{j=1}^{N}
(x_i - y_j)
\Delta^{-1}\{x\}_{n-1} \Delta^{-1}\{-y\}_N
\\
\times
\lim_{\epsilon \rightarrow 0}
\frac{1}{{\epsilon}^{N-n}}
\left|
\begin{array}{ccc}
\frac{1}{(x_1 - y_1)(x_1 - y_1 + 1)} 
& 
\cdots 
& 
\frac{1}{(x_1 - y_N)(x_1 - y_N + 1)}
\\
\vdots 
& & 
\vdots
\\
\frac{1}{(x_{n-1} - y_1)(x_{n-1} - y_1 + 1)}
&
\cdots
&
\frac{1}{(x_{n-1} - y_N)(x_{n-1} - y_N + 1)}
\\
\\
f^{(n)}_1 (\epsilon)
&
\cdots
&
f^{(n)}_N (\epsilon)
\\
h_{N-n-1}(y_1,\bary_1) & \cdots & h_{N-n-1}(y_N,\bary_N)
\\
\vdots & & \vdots
\\
h_0(y_1,\bary_1) & \cdots & h_0(y_N,\bary_N)
\end{array}
\right|
\end{multline}
Consider the functions $f^{(n)}_j (\epsilon)$ in the $n$-th row of 
the above determinant, which is defined in 
Equation ({\bf \ref{aux-funct}}). 
Since $\epsilon$ is small, they can be expanded in powers of $\epsilon$ 
using the definition of the elementary and complete symmetric functions 

\begin{align}
f^{(n)}_j(\epsilon)
=
\sum_{k=0}^{\infty}
\sum_{l=0}^{k}
\sum_{m=0}^{l}
(-)^{k-l}
\ e_{k-l}(y_1,\dots,y_N)
\ h_{l-m}(x_1,\dots,x_{n-1})
\ h_m(y_j,\bary_j)
\ {\epsilon}^k
\label{aux-series}
\end{align}
where $e_k(y_1,\dots,y_N)$ and $h_k(x_1,\dots,x_{n-1})$ are elementary 
and complete symmetric functions, given respectively by the generating 
functions

\begin{align}
\sum_{k=0}^{\infty}
e_k(y_1,\dots,y_N)
\epsilon^k
=
\prod_{k=1}^{N}
(1+y_k \epsilon),
\quad\quad
\sum_{k=0}^{\infty}
h_k(x_1,\dots,x_{n-1})
\epsilon^k
=
\prod_{k=1}^{n-1}
\frac{1}{(1-x_k \epsilon)}
\end{align}

\noindent Using the series expression in Equation ({\bf \ref{aux-series}}) 
for the row of entries $f^{(n)}_j(\epsilon)$, one can see that all terms 
in the first sum with $0 \leq k \leq (N-n-1)$ give no contribution to the 
determinant since they are linear combinations of the lower $(N-n)$ rows, 
and the first sum starts at $k=(N-n)$. Taking the limit
$\epsilon \to 0$, all higher order terms in this sum vanish.

Studying the $k=(N-n)$ term in the series in 
Equation ({\bf \ref{aux-series}}), it is clear that many of its sub-terms 
do not contribute to the determinant either. In fact, only the sub-term 
corresponding to $l=m=(N-n)$ survives, and this is identically 
$h_{N-n}(y_j,\bary_j)$. Therefore we obtain

\begin{multline}
\lim_{x_N,\dots,x_n \rightarrow \infty}
\ll
x_n \cdots x_N
Z_1 \ll \{x\}_N \Big| \{y\}_N \rr
\rr
=
\\
\frac{\displaystyle{
\prod_{i=1}^{n-1} 
\prod_{j=1}^{N}
(x_i - y_j)
}}
{\displaystyle{
\Delta\{x\}_{n-1} \Delta\{-y\}_N
}}
\left|
\begin{array}{ccc}
\frac{1}{(x_1 - y_1)(x_1 - y_1 + 1)} 
& 
\cdots 
& 
\frac{1}{(x_1 - y_N)(x_1 - y_N + 1)}
\\
\vdots 
& & 
\vdots
\\
\frac{1}{(x_{n-1} - y_1)(x_{n-1} - y_1 + 1)}
&
\cdots
&
\frac{1}{(x_{n-1} - y_N)(x_{n-1} - y_N + 1)}
\\
\\
h_{N-n}(y_1,\bary_1)
&
\cdots
&
h_{N-n}(y_N,\bary_N)
\\
h_{N-n-1}(y_1,\bary_1) & \cdots & h_{N-n-1}(y_N,\bary_N)
\\
\vdots & & \vdots
\\
h_0(y_1,\bary_1) & \cdots & h_0(y_N,\bary_N)
\end{array}
\right|
\end{multline}

\noindent which proves $\mathcal{P}_{N-n+1}$. This completes 
the proof of Equation ({\bf \ref{conjecture}}) for all 
$0 \leq n \leq N-1$, by induction.

\hfill$\square$ 

\subsection{A {\it \lq partial Vandermonde\rq} \ way to write 
the determinant}

A simple check shows that the highest order term in $h_{N-i}(y_j,\bary_j)$ 
is $(N-i+1) y_j^{N-i}$. Using row operations to cancel all terms of lower 
order and extracting an overall factor of $(N-n)!$ from the determinant 
in Equation ({\bf \ref{conjecture}}), we obtain

\begin{multline}
\label{result1}
Z_1 \ll \{x\}_n \Big| \{y\}_N \rr 
=
\frac{\displaystyle{
\prod_{i=1}^{n} 
\prod_{j=1}^{N}
(x_i - y_j)
}}
{\displaystyle{
\Delta\{x\}_n \Delta\{-y\}_N
}}
\left|
\begin{array}{ccc}
\frac{1}{(x_1 - y_1)(x_1 - y_1 + 1)} 
& 
\cdots 
& 
\frac{1}{(x_1 - y_N)(x_1 - y_N + 1)}
\\
\vdots 
& & 
\vdots
\\
\frac{1}{(x_{n} - y_1)(x_{n} - y_1 + 1)}
&
\cdots
&
\frac{1}{(x_{n} - y_N)(x_{n} - y_N + 1)}
\\
\\
y_1^{N-n-1} & \cdots & y_N^{N-n-1}
\\
\vdots & & \vdots
\\
y_1^{0} & \cdots & y_N^{0}
\end{array}
\right|
\end{multline}

\noindent Equation ({\bf\ref{result1}}) is our $(N\! \times \! N)$ 
determinant expression for the pDWPF. As previously mentioned, the 
top $n$ rows are of Izergin determinant-type, whereas the lower 
$(N-n)$ rows are of Vandermonde determinant-type.

\subsection{Towards the trigonometric pDWPF}
\label{lim-DWPF-trig}

Using the asymptotic behaviour of the trigonometric weights given 
in Equation ({\bf \ref{trig-lim}}), we can repeat the procedure 
of Subsection {\bf\ref{lim-DWPF-rat}} to derive the relation

\begin{align}
Z_1 \ll \{x\}_i \Big| \{y\}_N\rr
&\sim
(1+ \cdots + e^{-2\gamma(N-i)})
e^{-\gamma} [\gamma]
Z_1 \ll \{x\}_{i-1} \Big| \{y\}_N \rr
\\
&\sim
(1-e^{-2\gamma(N-i+1)})
Z_1 \ll \{x\}_{i-1} \Big| \{y\}_N \rr,
\quad\quad
\text{as } x_i \rightarrow \infty
\nonumber
\end{align}

\noindent between trigonometric partial domain wall partition 
functions. Iterating this result through $i=\{N,\dots,n+1\}$, 
we obtain 

\begin{align}
\label{trig-many-limits}
Z_1 \ll \{x\}_n \Big| \{y\}_N \rr
=
\frac{e^{(N-n) \gamma}}
{[\gamma]^{N-n} [N-n]_q!}
\lim_{x_N,\dots,x_{n+1} \rightarrow \infty}
Z_1 \ll \{x\}_N \Big| \{y\}_N \rr
\end{align}

\noindent where 

\begin{align}
q=e^{-2\gamma},
\quad\quad
[k]_q = \frac{1-q^k}{1-q},
\quad\quad
[j]_q! = \prod_{k=1}^{j} [k]_q
\end{align}

\noindent Equation ({\bf \ref{trig-many-limits}}) is 
the trigonometric analogue of 
Equation ({\bf \ref{many-limits}}).

\subsection{The trigonometric pDWPF as 
an $(N\! \times \! N)$-determinant}
Starting from Izergin's trigonometric determinant, 
Equation ({\bf\ref{IK-trig}}), and taking the limits in 
Equation ({\bf\ref{trig-many-limits}}), it is straightforward 
to show that
\begin{multline}
\label{result-trig}
Z_1 \ll \{x\}_n \Big| \{y\}_N \rr 
=
\\
[\gamma]^{n}
e^{(N-n+1)(|x|-|y|)} 
\frac{\displaystyle{
\prod_{i=1}^{n}
\prod_{j=1}^{N}
[x_i - y_j]
}}
{\displaystyle{
\Delta\{x\}_n \Delta\{-y\}_N
%\prod_{1 \leq i<j \leq n}
%[x_j - x_i]
%\prod_{1 \leq i<j \leq N}
%[y_i - y_j]
}}
\left|
\begin{array}{ccc}
\frac{1}{[x_1 - y_1][x_1 - y_1 + \gamma]} 
& 
\cdots 
& 
\frac{1}{[x_1 - y_N][x_1 - y_N + \gamma]}
\\
\vdots 
& & 
\vdots
\\
\frac{1}{[x_{n} - y_1][x_{n} - y_1 + \gamma]}
&
\cdots
&
\frac{1}{[x_{n} - y_N][x_{n} - y_N + \gamma]}
\\
\\
e^{2y_1(N-n)} & \cdots & e^{2y_N(N-n)}
\\
\vdots & & \vdots
\\
e^{2y_1} & \cdots & e^{2y_N}
\end{array}
\right|
\end{multline}

\noindent We omit the details since they are similar to those 
in  the rational case.

%%%SEC03%%%
\section{Slavnov scalar product in the infinite-rapidity limit}
\label{section-3}

In this section we obtain an alternative expression for the pDWPF, 
by starting from Slavnov's formula for the scalar product 
(Equation ({\bf \ref{betscal}})) and taking appropriate limits. 
The resulting expression is an $(n\! \times \! n)$ determinant.

\subsection{$n$ rapidities become infinite} 
\label{lim-sp}

Using the six-vertex model representation of the scalar product, 
Figure {\bf\ref{scalar-prod}}, it is possible calculate the partial 
domain wall partition function in the alternative way
\begin{align}
Z_2 \ll \{x\}_n \Big| \{y\}_N \rr
=
\frac{1}{n!}
\lim_{b_n,\dots,b_1 \rightarrow \infty}
\ll
b_1 \cdots b_n
S \ll \{x\}_n,\{b\}_n \Big| \{y\}_N \rr
\rr
\label{s-many-limits}
\end{align}
where the limits are sequentially, starting with $b_n$. The argument 
which underlies Equation ({\bf \ref{s-many-limits}}) is the same as 
the one that underlies Equation ({\bf \ref{many-limits}}). 

\subsection{The infinite rapidity limit of Slavnov's determinant}

To obtain an alternative determinant expression for the pDWPF, we 
start from Slavnov's determinant in Equation ({\bf\ref{betscal}}) 
and perform the limits specified in Equation ({\bf \ref{s-many-limits}}). 
We do this using induction, by proving the following result.

\begin{myLemma} For all $0 \leq m \leq n-1$, we have
\begin{multline}
\label{k-proof-1}
\frac{\displaystyle{
\lim_{b_n, \dots, b_{m+1} \rightarrow \infty}
\ll
b_{m+1} \dots b_n S\ll\{x\}_n, \{b\}_n \Big| \{y\}_N\rr
\rr
}}
{(n-m)!}
=
\frac{\displaystyle{
\prod_{i=1}^{n} \prod_{j=1}^{m}
(b_j-x_i-1)
}}
{\displaystyle{
\Delta\{-x\}_n \Delta\{b\}_m
%\prod_{1\leq i<j \leq n} (x_i-x_j)
%\prod_{1 \leq i<j \leq m} (b_j-b_i)
}}
\\
\times
\det
\ll
\begin{array}{c|c}
\underbrace{
\frac{\displaystyle{t_i^{(m)}}}{\displaystyle{(b_j-x_i)(b_j-x_i+1)}}
- 
\frac{\displaystyle{s_i}}{\displaystyle{(b_j-x_i)(b_j-x_i-1)}} 
}_{1 \leq j \leq m}
&
\underbrace{ 
x_i^{j-1} t_i^{(m)} - (x_i+1)^{j-1} s_i
}_{n-m \geq j \geq 1}
\end{array}
\rr_{1 \leq i \leq n }
\end{multline}

\noindent where 

\begin{align}
s_i = \prod_{k=1}^{N} 
\ll 
\frac{x_i-y_k}{x_i-y_k+1}
\rr,
\quad
t_i^{(m)} = \prod_{k=1}^{m} 
\ll
\frac{b_k-x_i+1}{b_k-x_i-1}
\rr
\end{align}

\end{myLemma}

\myProof
Let Equation ({\bf\ref{k-proof-1}}) be a proposition, 
$\mathcal{P}_{n-m}$. We begin with the proof of $\mathcal{P}_{1}$. 
Using Equation ({\bf\ref{betscal}}) for the scalar product, we have

\begin{multline}
\lim_{b_n \rightarrow \infty}
\ll b_n
S\ll \{x\}_n,\{b\}_n \Big| \{y\}_N\rr
\rr
=
\lim_{b_n \rightarrow \infty}
\frac{\displaystyle{b_n \prod_{i=1}^{n} \prod_{j=1}^{n} (b_j-x_i-1)}}
{\displaystyle{
\Delta\{-x\}_n \Delta\{b\}_n
%\prod_{1 \leq i < j \leq n} (x_i-x_j)(b_j-b_i)
}}
\\
\times
\det
\ll
\begin{array}{c}
\underbrace{
\frac{\displaystyle{t_i^{(n)}}}{\displaystyle{(b_j-x_i)(b_j-x_i+1)}}
-
\frac{\displaystyle{s_i}}{\displaystyle{(b_j-x_i)(b_j-x_i-1)}}
}_{1 \leq j \leq n}
\end{array}
\rr_{1\leq i \leq n}
\label{k-proof-2}
\end{multline}
In the limit being considered, we have $t_i^{(n)} \rightarrow t_i^{(n-1)}$. 
The only other place where the determinant in Equation ({\bf\ref{k-proof-2}}) 
depends on $b_n$ is in its final column, and we may absorb all terms in 
the prefactor of Equation ({\bf\ref{k-proof-2}}) which depend on $b_n$ into 
the final column of the determinant, and take the limit easily. The result 
is

\begin{multline}
\lim_{b_n \rightarrow \infty}
\ll b_n
S \ll \{x\}_n,\{b\}_n \Big| \{y\}_N \rr
\rr
=
\frac{\displaystyle{\prod_{i=1}^{n} \prod_{j=1}^{n-1} (b_j-x_i-1)}}
{\displaystyle{
\Delta\{-x\}_n \Delta\{b\}_{n-1}
%\prod_{1 \leq i < j \leq n} (x_i-x_j) 
%\prod_{1 \leq i<j \leq n-1} (b_j-b_i)
}}
\\
\times
\det
\ll
\begin{array}{c|c}
\underbrace{
\frac{\displaystyle{t_i^{(n-1)}}}{\displaystyle{(b_j-x_i)(b_j-x_i+1)}}
-
\frac{\displaystyle{s_i}}{\displaystyle{(b_j-x_i)(b_j-x_i-1)}}
}_{1 \leq j \leq n-1}
&
t_i^{(n-1)}-s_i
\end{array}
\rr_{1\leq i \leq n}
\end{multline}
which proves $\mathcal{P}_{1}$. Now we assume that $\mathcal{P}_{n-m}$ 
is true and show that this implies $\mathcal{P}_{n-m+1}$. Taking 
Equation ({\bf \ref{k-proof-1}}) as our starting point, we find that

\begin{multline}
\label{k-proof-3}
\frac{\displaystyle{
\lim_{b_n,\dots, b_m \rightarrow \infty}
\ll
b_m \dots b_n S\ll \{x\}_n,\{b\}_n \Big| \{y\}_N \rr
\rr
}}
{(n-m+1)!}
=
\frac{\displaystyle{
\lim_{b_m \rightarrow \infty}
b_m
\prod_{i=1}^{n} \prod_{j=1}^{m}
(b_j-x_i-1)
}}
{\displaystyle{
(n-m+1)
\Delta\{-x\}_n \Delta\{b\}_m
%\prod_{1\leq i<j \leq n} (x_i-x_j)
%\prod_{1 \leq i<j \leq m} (b_j-b_i)
}}
\\
\times
\det
\ll
\begin{array}{c|c}
\underbrace{
\frac{\displaystyle{t_i^{(m)}}}{\displaystyle{(b_j-x_i)(b_j-x_i+1)}}
-
\frac{\displaystyle{s_i}}{\displaystyle{(b_j-x_i)(b_j-x_i-1)}} 
}_{1 \leq j \leq m}
&
\underbrace{ 
x_i^{j-1} t_i^{(m)} - (\barx_i)^{j-1} s_i
}_{n-m \geq j \geq 1}
\end{array}
\rr_{1 \leq i \leq n }
\end{multline}
where we define as usual $\barx_i = x_i+1$. Let $1/b_m = \epsilon$, 
where $\epsilon$ is small in view of the limit being taken. Then we 
can write 

\begin{align}
\frac{b_m}{(b_m-x_i)(b_m-x_i+1)}
&=
\sum_{k=0}^{\infty}
h_k(x_i, x_i-1)
\epsilon^{k+1}
\label{gs1}
\\
\frac{b_m}{(b_m - x_i)(b_m - x_i - 1)}
&=
\sum_{k=0}^{\infty}
h_k(\barx_i,\barx_i-1)
\epsilon^{k+1}
\label{gs2}
\end{align}
These expressions can be substituted into the $m$-th column of 
the determinant in Equation ({\bf\ref{k-proof-3}}), thereby 
cancelling all terms in the sums in 
Equations ({\bf\ref{gs1}}), ({\bf\ref{gs2}}) of degree 
$\leq (n-m)$ in $\epsilon$, since these are linear combinations 
of the last $(n-m)$ columns of the determinant, and we obtain

\begin{multline}
\frac{\displaystyle{
\lim_{b_n,\dots,b_m \rightarrow \infty}
\ll
b_m \dots b_n S\ll \{x\}_n,\{b\}_n \Big| \{y\}_N \rr
\rr
}}
{(n-m+1)!}
=
\frac{\displaystyle{
\lim_{b_m \rightarrow \infty}
\prod_{i=1}^{n} \prod_{j=1}^{m}
(b_j-x_i-1)
}}
{\displaystyle{
(n-m+1)
\Delta\{-x\}_n \Delta\{b\}_m
%\prod_{1\leq i<j \leq n} (x_i-x_j)
%\prod_{1 \leq i<j \leq m} (b_j-b_i)
}}
\times
\\
\left|
\begin{array}{c|c|c}
\underbrace{
\cdots
}_{1 \leq j \leq m-1}
&
\displaystyle{
\sum_{k=n-m}^{\infty}
}
\left(
h_k(x_i,x_i-1)
t_i^{(m)}
-
h_k(\barx_i,\barx_i-1)
s_i
\right)
b_m^{-k-1} 
&
\underbrace{ 
x_i^{j-1} t_i^{(m)} - (\barx_i)^{j-1} s_i
}_{n-m \geq j \geq 1}
\end{array}
\right|
\end{multline}
where we abbreviate $\det(\cdot)_{1\leq i \leq n}$ by $|\cdot|$, and 
where the first $(m-1)$ columns are as before, so we do not 
write them. Now we can move all terms in the prefactor which 
depend on $b_m$ inside the $m$-th column, and take the limit. 
This gives us

\begin{multline}
\frac{\displaystyle{
\lim_{b_n,\dots, b_m \rightarrow \infty}
\ll
b_m \dots b_n S\ll \{x\}_n,\{b\}_n \Big| \{y\}_N\rr
\rr
}}
{(n-m+1)!}
=
\frac{\displaystyle{
\prod_{i=1}^{n} \prod_{j=1}^{m-1}
(b_j-x_i-1)
}}
{\displaystyle{
(n-m+1)
\Delta\{-x\}_n \Delta\{b\}_{m-1}
%\prod_{1\leq i<j \leq n} (x_i-x_j)
%\prod_{1 \leq i<j \leq m-1} (b_j-b_i)
}}
\times
\\
\left|
\begin{array}{c|c|c}
\underbrace{
\cdots
}_{1 \leq j \leq m-1}
&
h_{n-m}(x_i,x_i-1)
t_i^{(m-1)}
-
h_{n-m}(\barx_i,\barx_i-1)
s_i
&
\underbrace{ 
x_i^{j-1} t_i^{(m-1)} - (\barx_i)^{j-1} s_i
}_{n-m \geq j \geq 1}
\end{array}
\right|
\end{multline}
where we have again abbreviated $\det(\cdot)_{1\leq i \leq n}$ by $|\cdot|$.
Finally, we observe that the highest order terms in $h_{n-m}(x_i,x_i-1)$ 
and $h_{n-m}(\barx_i,\barx_i-1)$ are 
$(n-m+1) x_i^{n-m}$ and $(n-m+1) \barx_i^{n-m}$ respectively, while all 
other terms cancel with the last $(n-m)$ columns of the determinant, 
therefore
\begin{multline}
\frac{\displaystyle{
\lim_{b_n,\dots, b_m \rightarrow \infty}
\ll
b_m \dots b_n S\ll\{x\}_n,\{b\}_n \Big| \{y\}_N\rr
\rr
}}
{(n-m+1)!}
=
\frac{\displaystyle{
\prod_{i=1}^{n} \prod_{j=1}^{m-1}
(b_j-x_i-1)
}}
{\displaystyle{
\Delta\{-x\}_n \Delta\{b\}_{m-1}
%\prod_{1\leq i<j \leq n} (x_i-x_j)
%\prod_{1 \leq i<j \leq m-1} (b_j-b_i)
}}
\\
\times
\det
\ll
\begin{array}{c|c}
\underbrace{
\frac{\displaystyle{t_i^{(m-1)}}}{\displaystyle{(b_j-x_i)(b_j-x_i+1)}}
-
\frac{\displaystyle{s_i}}{\displaystyle{(b_j-x_i)(b_j-x_i-1)}} 
}_{1 \leq j \leq m-1}
&
\underbrace{ 
x_i^{j-1} t_i^{(m-1)} - (\barx_i)^{j-1} s_i
}_{n-m+1 \geq j \geq 1}
\end{array}
\rr_{1 \leq i \leq n }
\end{multline}
thus $\mathcal{P}_{n-m+1}$ is true. This proves 
Equation ({\bf\ref{k-proof-1}}) for all $0 \leq m \leq n-1$, by induction.

\hfill$\square$  

\subsection{Kostov's determinant}
For $m=0$, Equation ({\bf\ref{k-proof-1}}) leads to 

\begin{multline}
\label{k-proof-4}
\displaystyle{\frac{1}{n!}}
\ \ 
\displaystyle{
\lim_{b_n, \dots, b_{1} \rightarrow \infty}
\ll
b_{1} \dots b_n S \ll \{x\}_n, \{b\}_n \Big| \{y\}_N \rr
\rr
}
=
\\
\displaystyle{ 
\Delta^{-1}\{-x\}_n 
}
\displaystyle{
\det
\ll
\begin{array}{c}
\underbrace{ 
x_i^{j-1} - (x_i+1)^{j-1} s_i
}_{n \geq j \geq 1}
\end{array}
\rr_{1 \leq i \leq n }
}
\end{multline}
where we have used $t^{(0)}_i \equiv 1$. Combining 
Equations ({\bf\ref{s-many-limits}}) and ({\bf\ref{k-proof-4}}), 
we see that the partial domain wall partition function can be written as
\begin{align}
Z_2\ll\{x\}_n \Big| \{y\}_N\rr
=
\Delta^{-1}\{x\}_n
\det
\ll
x_i^{j-1} 
-
(x_i+1)^{j-1}
\prod_{k=1}^{N}
\frac{
(x_i - y_k)
}
{
(x_i - y_k + 1)
}
\rr_{1 \leq i,j \leq n}
\label{result2}
\end{align}
where we have simultaneously reversed the order of variables in 
the Vandermonde and the order of columns in the determinant. In 
contrast to the determinant in Equation ({\bf\ref{result1}}), 
which is $(N \! \times \! N)$, the determinant in 
Equation ({\bf\ref{result2}}) is $(n \! \times \! n)$.
Equation ({\bf \ref{result2}}) is due to I Kostov 
\cite{kostov.private.communication, 
      kostov.short.paper, 
      kostov.long.paper}.

\subsection{Writing $Z_2( \{x\}_n | \{y\}_N )$ as a sum over 
partitions}
To conclude the section, we show that the determinant in 
Equation ({\bf\ref{result2}}) can be expanded as a certain 
sum, which is the precise form in which it appears in \cite{GSV}. 
Let us define the functions

\begin{align}
X_j(x_i)
=
x_i^{j-1},
\quad\quad
Y_j(x_i)
=
\barx_i^{j-1} 
\prod_{k=1}^{N} 
\ll
\frac{x_i - y_k}{\barx_i - y_k}
\rr
\end{align}
Then we have
\begin{align}
Z_2 \ll \{x\}_n \Big| \{y\}_N \rr
=
%\frac{
\Delta^{-1}\{x\}_n
\det
\ll
X_j(x_i) - Y_j(x_i)
\rr_{1\leq i,j \leq n}
%}
%{\displaystyle{
%\prod_{1 \leq i < j \leq n}
%(x_j-x_i)
%}}
\label{X-Y}
\end{align}
Using Laplace's formula for the determinant of a sum of two matrices, 
we write the expression in Equation ({\bf\ref{X-Y}}) as a sum over all 
partitions of the integers 
$\{1,\dots,n\}$ into disjoint sets 
$\{\alpha\}_{n-m} = \{\alpha_1 < \cdots < \alpha_{n-m}\}$, 
$\{\beta\}_{m} = \{\beta_1 < \cdots < \beta_{m}\}$. The result 
is

\begin{multline}
Z_2 \ll \{x\}_n \Big| \{y\}_N \rr
=
\\
\Delta^{-1} \{x\}_n
\sum_{\substack{\{1,\dots,n\} = \\ \{\alpha\}_{n-m} \cup \{\beta\}_{m} }}
(-)^{{\rm sgn}(P)+m}
\left|
\begin{array}{lcl}
X_1(x_{\alpha_1})
& 
\cdots
&
X_n(x_{\alpha_1})
\\
\vdots
&
&
\vdots
\\
X_1(x_{\alpha_{n-m}})
& 
\cdots
&
X_n(x_{\alpha_{n-m}})
\\
Y_1(x_{\beta_1})
& 
\cdots
&
Y_n(x_{\beta_1})
\\
\vdots
&
&
\vdots
\\
Y_1(x_{\beta_{m}})
& 
\cdots
&
Y_n(x_{\beta_{m}})
\end{array}
\right|
\label{partitions1}
\end{multline}
where we have abbreviated 
$\prod_{1 \leq i < j \leq n} (x_j - x_i) = \Delta\{x\}_n$ and ${\rm sgn}(P)$
 denotes the sign of the permutation 
$P\{1,\dots, n\} = \{\alpha_1,\dots,\alpha_{n-m},\beta_1,\dots,\beta_{m}\}$. 
It is possible to extract common factors from the determinant in 
the sum in Equation ({\bf \ref{partitions1}}), and write it as

\begin{multline}
\left|
\begin{array}{lcl}
X_1(x_{\alpha_1})
&
\cdots
&
X_n(x_{\alpha_1})
\\
\vdots
&
&
\vdots
\\
X_1(x_{\alpha_{n-m}})
&
\cdots
&
X_n(x_{\alpha_{n-m}})
\\
Y_1(x_{\beta_1})
&
\cdots
&
Y_n(x_{\beta_1})
\\
\vdots
&
&
\vdots
\\
Y_1(x_{\beta_{m}})
&
\cdots
&
Y_n(x_{\beta_{m}})
\end{array}
\right|
=
\prod_{i=1}^{m}
\prod_{k=1}^{N}
\ll
\frac{x_{\beta_i} - y_k}{\barx_{\beta_i} - y_k}
\rr
\left|
\begin{array}{lcl}
x_{\alpha_1}^{0}
&
\cdots
&
x_{\alpha_1}^{n-1}
\\
\vdots
&
&
\vdots
\\
x_{\alpha_{n-m}}^{0}
&
\cdots
&
x_{\alpha_{n-m}}^{n-1}
\\
\barx_{\beta_1}^{0}
&
\cdots
&
\barx_{\beta_1}^{n-1}
\\
\vdots
&
&
\vdots
\\
\barx_{\beta_{m}}^{0}
&
\cdots
&
\barx_{\beta_{m}}^{n-1}
\end{array}
\right|
\\
=
\prod_{i=1}^{m}
\prod_{k=1}^{N}
\frac{(x_{\beta_i} - y_k)}{(\barx_{\beta_i} - y_k)}
\Delta\{x_{\alpha}\}_{n-m}
\prod_{i=1}^{n-m} \prod_{j=1}^{m}
(\barx_{\beta_j} - x_{\alpha_i})
\Delta\{\barx_{\beta}\}_{m}
\nonumber
\end{multline}
where 
$\prod_{1 \leq i < j \leq n-m} (x_{\alpha_j} - x_{\alpha_i}) 
= \Delta\{x_{\alpha}\}_{n-m}$ and
$\prod_{1 \leq i < j \leq m} (\barx_{\beta_j} - \barx_{\beta_i}) 
= 
\Delta\{\barx_{\beta}\}_{m} = \Delta\{x_{\beta}\}_m $.
Putting this expression back into Equation ({\bf\ref{partitions1}}), 
we get

\begin{multline}
Z_2 \ll \{x\}_n \Big| \{y\}_N \rr
=
\\
\sum_{\{1,\dots,n\} = \{\alpha\}_{n-m} \cup \{\beta\}_{m} }
(-)^{m}
\prod_{i=1}^{m}
\prod_{k=1}^{N}
\ll
\frac{x_{\beta_i} - y_k}{\barx_{\beta_i} - y_k}
\rr
\ll
\prod_{i=1}^{n-m} \prod_{j=1}^{m}
\frac{x_{\alpha_i} - \barx_{\beta_j}}
{x_{\alpha_i} - x_{\beta_j}}
\rr
\end{multline}

\noindent Up to simple changes in variables, this is the sum 
expression in \cite{GSV}.

\subsection{The trigonometric version of Kostov's determinant}
For completeness, we give the trigonometric version of the pDWPF 
obtained taking limits of Slavnov's scalar product in 
Equation ({\bf\ref{betscal2}}). The starting point in 
the calculation is the relation

\begin{align}
\label{trig-many-limits-sp}
Z_2 \ll \{x\}_n \Big| \{y\}_N \rr
=
\frac{e^{n \gamma}}
{[\gamma]^{n} [n]_q!}
\lim_{b_n,\dots,b_{1} \rightarrow \infty}
S \ll \{x\}_n, \{b\}_n \Big| \{y\}_N \rr
\end{align}

\noindent which arises from the lattice version of the scalar 
product in Figure {\bf\ref{scalar-prod}} and that of the pDWPF 
on the right of Figure {\bf\ref{dwpf-res}}, and the asymptotic 
behaviour of the weights in Equation ({\bf\ref{trig-lim}}).

Repeating the ideas already developed in this section and working 
from Slavnov's determinant in Equation ({\bf\ref{betscal2}}), 
Equation ({\bf\ref{trig-many-limits-sp}}) ultimately leads us 
to the expression

\begin{multline}
Z_2 \ll \{x\}_n \Big| \{y\}_N \rr
=
e^{-(n-1)|x|}
\Delta^{-1}\{x\}_n
\\
\times
\det
\ll
e^{2x_i(j-1)}
\ll
1-e^{N\gamma}
\prod_{k=1}^{N}
\frac{[x_i-y_k]}
{[x_i-y_k+\gamma]}
e^{-2\gamma(n-j)}
\rr
\rr_{1\leq i,j \leq n}
\end{multline}

\noindent for the trigonometric pDWPF.

%%%SEC04%%%
\section{Equivalence of determinants}
\label{section-4}

In this section we show directly that the determinants 
in Equations ({\bf \ref{result1}}) and ({\bf \ref{result2}}) are equal. 
We do this for completeness and as a check of our limit calculations, 
because on the surface it is not apparent that the two expressions 
coincide. We restrict our attention to the determinants obtained from 
the rational six-vertex model, because as we have already mentioned 
the two lattice sums in Figure {\bf\ref{dwpf-res}} are {\it not} 
equivalent in the trigonometric parametrization. 

Our approach is to convert both Equations ({\bf \ref{result1}}) 
and ({\bf \ref{result2}}) to polynomials, by multiplying them 
by an overall factor (precisely the factor present in the denominator 
of the $b$ and $c$ weights). We distinguish the resulting expressions 
by calling them $\zi ( \{x\}_n | \{y\}_N )$ and
$\zs ( \{x\}_n | \{y\}_N )$, in reference to the size of the determinants 
in question, and show that both 
$\zi ( \{x\}_n | \{y\}_N )$ and 
$\zs ( \{x\}_n | \{y\}_N )$ 
satisfy a list of conditions which the pDWPF must itself obey. 
This proves that they are equal, since the conditions only admit 
a unique solution.

\subsection{Set of properties which characterize the pDWPF}

Let $Z ( \{x\}_n | \{y\}_N )$ be the partition function of either of 
the lattices in Figure {\bf \ref{dwpf-res}}, but whose weights are 
given by

\begin{align}
a_{\pm}(x,y) = x - y + 1,
\quad\quad
b_{\pm}(x,y) = x - y,
\quad\quad
c_{\pm}(x,y) = 1
\end{align}
This polynomial version of $Z ( \{x\}_n | \{y\}_N )$ is obtained 
from the rational version by multiplying by 
$\prod_{i=1}^{n} \prod_{j=1}^{N} (x_i - y_j + 1)$. Following 
Korepin \cite{korepin.paper}, one can show that
\\

{\bf A.}
$Z ( \{x\}_n | \{y\}_N )$ is a polynomial in $x_n$ of degree 
bounded by $(2N-1)$.
\\

{\bf B.}
$Z ( \{x\}_n | \{y\}_N )$ is symmetric in the set of variables 
$\{y\}_N = \{y_1,\dots,y_N\}$.
\\

{\bf C.}
For all $n \geq 2$, $Z ( \{x\}_n | \{y\}_N )$ satisfies 

\begin{align}
\label{rec1}
Z\ll \{x\}_n \Big| \{y\}_N\rr
\Big|_{x_n = y_N}
&=
\prod_{k=1}^{N-1}
(y_N - y_k +1)
\prod_{k=1}^{n-1}
(x_k - \bary_N)
Z \ll \{x\}_{n-1} \Big| \{y\}_{N-1} \rr
\\
Z \ll \{x\}_n \Big| \{y\}_N \rr
\Big|_{x_n = \bary_N}
&=
\prod_{k=1}^{N-1}
(y_N - y_k - 1)
\prod_{k=1}^{n-1}
(x_k - y_N)
Z \ll \{x\}_{n-1} \Big| \{y\}_{N-1} \rr
\label{rec2}
\end{align}

where $\bar{y}_N = y_N-1$.
\\

{\bf D.}
$Z ( x_1 | \{y\}_N )$ is known explicitly for all $N \geq 1$, 
and is given by

\begin{multline}
Z \ll x_1 \Big| \{y\}_N \rr
=
\\
\sum_{l=1}^{N}
\prod_{1 \leq k < l}
(x_1 - y_k)
\prod_{l < k \leq N}
(x_1 - y_k + 1)
=
\prod_{k=1}^{N}
(x_1 - y_k + 1)
-
\prod_{k=1}^{N}
(x_1 - y_k)
\label{initial-cond}
\end{multline}

\noindent The second equality in Equation ({\bf \ref{initial-cond}}) 
evaluates the sum over $l$, and can be established by induction on $N$.

These four properties uniquely determine the functions 
$Z ( \{x\}_n | \{y\}_N )$, for all $1 \leq n \leq N$. This is because, 
from {\bf A}, $Z ( \{x\}_n | \{y\}_N )$ is a polynomial in $x_n$, and 
from {\bf B} and {\bf C}, it is known at more points than its degree. 

\subsection{$\zi ( \{x\}_n | \{y\}_N )$ satisfies the four properties}
Consider the polynomial version of the pDWPF in 
Equation ({\bf \ref{result1}}), obtained by multiplying by 
$\prod_{i=1}^{n} \prod_{j=1}^{N} (x_i-y_j+1)$.  
Let us denote this by $\zi ( \{x\}_n | \{y\}_N )$. 
Using $\bary = y - 1$, it is given by

\begin{multline}
\zi \ll \{x\}_n \Big| \{y\}_N \rr
\equiv
\mathcal{P}(\{x\}_n,\{y\}_N)
\det 
\mathcal{M} \ll \{x\}_n,\{y\}_N \rr 
\\
=
\frac{\displaystyle{
\prod_{i=1}^{n} 
\prod_{j=1}^{N}
(x_i - y_j)(x_i - \bary_j)
}}
{\displaystyle{
\Delta\{x\}_n \Delta\{-y\}_N
%\prod_{1 \leq i<j \leq n}
%(x_j - x_i)
%\prod_{1 \leq i<j \leq N}
%(y_i - y_j)
}}
\left|
\begin{array}{ccc}
\frac{1}{(x_1 - y_1)(x_1 - \bary_1 )}
& 
\cdots
& 
\frac{1}{(x_1 - y_N)(x_1 - \bary_N )}
\\
\vdots
&
& 
\vdots
\\
\frac{1}{(x_{n} - y_1)(x_{n} - \bary_1 )}
&
\cdots
&
\frac{1}{(x_{n} - y_N)(x_{n} - \bary_N )}
\\
\\
y_1^{N-n-1} & \cdots & y_N^{N-n-1}
\\
\vdots
&
& 
\vdots
\\
y_1^{0} & \cdots & y_N^{0}
\end{array}
\right|
\label{poly1}
\end{multline}
Due to the cancellation of the Vandermonde $\prod_{1\leq i<j \leq n}
(x_j-x_i)$ with trivial zeros of the determinant,
$\zi ( \{x\}_n | \{y\}_N )$ is a polynomial in $x_n$ and the
highest degree it can obtain in this variable is $(2N-n-1)$.
Therefore property {\bf A} is satisfied. Interchanging the
variables $y_i$ and $y_j$ leaves $\zi(\{x\}_n | \{y\}_N)$ 
invariant, therefore property {\bf B} is satisfied. 
Setting $x_n = \{y_N, \bary_N\}$, we see that only the minor

\begin{align}
\det
\mathcal{M}\ll
\{x\}_{n-1},\{y\}_{N-1}
\rr
=
\left|
\begin{array}{ccc}
\frac{1}{(x_1 - y_1)(x_1 - \bary_1 )} 
& 
\cdots 
& 
\frac{1}{(x_1 - y_{N-1})(x_1 - \bary_{N-1} )}
\\
\vdots 
& & 
\vdots
\\
\frac{1}{(x_{n-1} - y_1)(x_{n-1} - \bary_1 )}
&
\cdots
&
\frac{1}{(x_{n-1} - y_{N-1})(x_{n-1} - \bary_{N-1} )}
\\
\\
y_1^{N-n-1} & \cdots & y_{N-1}^{N-n-1}
\\
\vdots & & \vdots
\\
y_1^{0} & \cdots & y_{N-1}^{0}
\end{array}
\right|
\end{align}
survives in the Laplace expansion of 
$\det \cM (\{x\}_n,\{y\}_N)$ 
down the right-most column. The prefactor 
$\cP (\{x\}_n, \{y\}_N)$ 
in Equation ({\bf \ref{poly1}}) satisfies

\begin{multline}
\frac{
(-)^{n+N}
\mathcal{P}(\{x\}_n,\{y\}_N)
}
{
(x_n-y_N)(x_n-\bary_N)
}
=
\\
\left\{
\begin{array}{ll}
\displaystyle{
\prod_{i=1}^{n-1}
(x_i-\bary_N)
\prod_{j=1}^{N-1}
(y_N-y_j+1)
\ 
\cP (\{x\}_{n-1},\{y\}_{N-1}),}
&
\quad
x_n = y_N
\\
\displaystyle{
\prod_{i=1}^{n-1}
(x_i-y_N)
\prod_{j=1}^{N-1}
(y_N-y_j-1)
\ 
\cP (\{x\}_{n-1},\{y\}_{N-1}),}
&
\quad
x_n = \bary_N
\end{array}
\right.
\end{multline}
Combining results, we see that property {\bf C} is satisfied. 
Finally, when $n=1$, we have

\begin{multline}
\zi \ll x_1\Big| \{y\}_N \rr
=
\\
\displaystyle{\Delta^{-1}\{-y\}_N}
\displaystyle{
\prod_{j=1}^{N}
(x_1-y_j)(x_1-\bary_j)
}
\left|
\begin{array}{ccc}
\frac{1}{(x_1 - y_1)(x_1 - \bary_1 )} 
& 
\cdots 
& 
\frac{1}{(x_1 - y_N)(x_1 - \bary_N )}
\\
\\
y_1^{N-2} & \cdots & y_N^{N-2}
\\
\vdots & & \vdots
\\
y_1^{0} & \cdots & y_N^{0}
\end{array}
\right|
\end{multline}
Laplace expanding this determinant along the first row and using 
the Vandermonde determinant identity, we obtain
\begin{align}
\zi \ll x_1\Big| \{y\}_N \rr
=
\sum_{j=1}^{N}
\prod_{k \not= j}^{N}
\frac{
(x_1 - y_k)(x_1 - \bary_k )
}
{
(y_j - y_k)
}
\end{align}
Comparing this polynomial in $x_1$ with the polynomial in 
Equation ({\bf \ref{initial-cond}}) at the points 
$x_1 = \{y_1,\bary_1,\dots,y_N,\bary_N\}$, we find that they 
are equal, and property {\bf D} is satisfied.

\subsection{$\zs ( \{x\}_n | \{y\}_N )$ satisfies the four 
properties}
Consider the polynomial version of the pDWPF in 
Equation ({\bf \ref{result2}}), obtained by multiplying the expression 
in that equation by $\prod_{i=1}^{n} \prod_{j=1}^{N} (x_i-y_j+1)$. We 
denote this by $\zs ( \{x\}_n | \{y\}_N )$. Using 
$\barx = x+1,\ \bary = y-1$, it is given by

\begin{multline}
\zs \ll \{x\}_n \Big| \{y\}_N \rr
=
\\
\displaystyle{\Delta^{-1}\{x\}_n}
\ 
\det
\ll
x_i^{j-1}
\prod_{k=1}^{N}
(x_i - \bary_k) 
-
(\barx_i)^{j-1}
\prod_{k=1}^{N}
(x_i - y_k)
\rr_{1 \leq i,j \leq n}
\\
\equiv
\displaystyle{\Delta^{-1}\{x\}_n} 
\ 
\det
\ll
\mathcal{M}_{j}(x_i,\{y\}_N)
\rr_{1 \leq i,j \leq n}
\label{poly2}
\end{multline}
Clearly $\zs ( \{x\}_n | \{y\}_N )$ is a polynomial in $x_n$, and 
the highest possible degree it can obtain in this variable is $N$. 
So property {\bf A} is satisfied. Since all $\{y\}_N$ dependence 
is in the products $\prod_{k=1}^{N}(x_i-\bary_k)$ and 
$\prod_{k=1}^{N}(x_i-y_k)$, property {\bf B} is satisfied. 
Setting $x_n = \{y_N, \bary_N\}$, the entries of the final row of 
the determinant in Equation ({\bf \ref{poly2}}) become

\begin{align}
\mathcal{M}_{j}
\ll
x_n,\{y\}_N
\rr
=
\left\{
\begin{array}{ll}
\displaystyle{
y_N^{j-1}
\prod_{k=1}^{N-1}
(y_N-y_k+1),
}
& 
\quad
x_n = y_N
\\
\displaystyle{
y_N^{j-1}
\prod_{k=1}^{N-1}
(y_N - y_k - 1),
}
&
\quad
x_n = \bary_N
\end{array}
\right.
\end{align}
Rearranging the entries of the top $(n-1)$ rows to write them as

\begin{multline}
\mathcal{M}_{j}\ll x_i, \{y\}_N\rr
=
\\
-
\frac{(x_i-y_N)(x_i-\bary_N)}
{y_N}
\ll
\frac{\displaystyle{
x_i^{j-1} \prod_{k=1}^{N-1} (x_i - \bary_k)
}}
{(1-x_i/y_N)}
-
\frac{\displaystyle{
(\barx_i)^{j-1} \prod_{k=1}^{N-1} (x_i - y_k)
}}
{(1-\barx_i/y_N)}
\rr
\end{multline}
Extracting factors which are common to each row of the determinant, 
we obtain

\begin{multline}
\label{intermediate-step}
\zs \ll \{x\}_n \Big| \{y\}_N \rr
=
\\
\Delta^{-1}\{x\}_{n-1}
\det
\ll
\begin{array}{c}
\mathcal{N}_{j}(x_i,\{y\}_N)
\\
\mathcal{N}_j(y_N)
\end{array}
\rr_{\substack{1 \leq i < n \\ 1 \leq j \leq n}}
\times
\left\{
\begin{array}{ll}
\displaystyle{
\prod_{i=1}^{n-1}
(x_i-\bary_N)
\prod_{k=1}^{N-1}
(y_N-y_k+1)
},
&
\quad
x_n = y_N
\\
\displaystyle{
\prod_{i=1}^{n-1}
(x_i-y_N)
\prod_{k=1}^{N-1}
(y_N-y_k-1)
},
&
\quad
x_n = \bary_N
\end{array}
\right.
\end{multline}
where we have defined the matrix entries
\begin{align}
\mathcal{N}_{j}
\ll
x_i,\{y\}_N
\rr
=
\frac{\displaystyle{
x_i^{j-1} \prod_{k=1}^{N-1} (x_i - \bary_k)
}}
{(1-x_i/y_N)}
-
\frac{\displaystyle{
(\barx_i)^{j-1} \prod_{k=1}^{N-1} (x_i - y_k)
}}
{(1-\barx_i/y_N)},
\quad\quad
\mathcal{N}_{j}(y_N)
=
y_N^{j-n}
\end{align}
Subtracting $(\text{column } j+1)/y_N$ from $(\text{column } j)$ 
for all $1 \leq j < n$, it is easy to show that

\begin{multline}
\det
\ll
\begin{array}{c}
\mathcal{N}_{j}(x_i,\{y\}_N)
\\
\mathcal{N}_j(y_N)
\end{array}
\rr_{\substack{1 \leq i < n \\ 1 \leq j \leq n}}
=
\det
\ll
\begin{array}{cc}
\mathcal{M}_{j}(x_i,\{y\}_{N-1})
& 
\mathcal{N}_{n}(x_i,\{y\}_N)
\\  
0
&
1
\end{array}
\rr_{1 \leq i,j \leq n-1}
\\
=
\det
\ll
\mathcal{M}_{j}(x_i,\{y\}_{N-1})
\rr_{1 \leq i,j \leq n-1}
\end{multline}
Substituting this into Equation ({\bf\ref{intermediate-step}}), 
we have verified that property {\bf C} is satisfied. Finally, 
when $n=1$, observe that $\zs ( x_1 | \{y\}_N )$ is identically 
equal to the right hand side of Equation ({\bf \ref{initial-cond}}), 
so property {\bf D} is satisfied.

\subsection{$\zi ( \{x\}_n |$$ \{y\}_N )$ and 
            $\zs ( \{x\}_n |$$ \{y\}_N )$ are equal}

Since $\zi ( \{x\}_n | \{y\}_N )$ and 
                  $\zs ( \{x\}_n | \{y\}_N )$ 
satisfy the properties {\bf A--D}, which admit a unique solution, 
we have proved that

\begin{align}
\zi \ll \{x\}_n \Big| \{y\}_N \rr
=
\zs \ll \{x\}_n \Big| \{y\}_N \rr,
\quad
\text{for all }
1 \leq n \leq N
\end{align}

%%%SEC05%%%
\section{Casorati determinants and discrete KP hierarchy}
\label{section-5}

\subsection{Notation related to sets of variables} 
In this section we use $\{x\}$ for a set of finitely many variables, 
and $\{\widehat{x}_m\}$ for $\{x\}$ with the element $x_m$ omitted. 
If a variable $x_i$ is repeated $m_i$ times, we use the superscript 
$(m_i)$ to indicate the multiplicity of $x_i$. For example,
$\{x_1^{(1)}, x_2^{(3)}, x_3^{(2)}, x_4^{(1)}, \dots \}$ 
is the same as
$\{x_1, x_2, x_2, x_2, x_3, x_3, x_4, \dots \}$
and $f\{x_{i}^{(m_i)}\}$ indicates that $f$ depends on $m_i$ distinct 
variables all of which are set to the same value $x_i$. Often we write 
$x_i$ instead of $x_i^{(1)}$. 

\subsection{The complete symmetric function $h_i\{x\}$} 
Let $\{x\}$ denote the set $\{x_1, $ $\dots, $ $x_N\}$. The complete 
symmetric function $h_i\{x\}$ is the coefficient of $k^i$ in the power 
series expansion 

\begin{align}
\prod_{i=1}^{N}
\frac{1}{1-x_i \, k}
=
\sum_{i=0}^{\infty}
h_i\{x\}
\, 
k^i
\label{complete-symmetric}
\end{align}
For example, $h_0\{x\} = 1$, $h_1(x_1,x_2) = x_1+x_2$,
$h_2(x_1,x_2,x_3) = x_1^2+x_2^2+x_3^2+x_1 x_2+x_1 x_3 + x_2 x_3$. 
By definition, $h_{i}\{x\} = 0$ for $i<0$. 

\subsection{Useful identities for $h_i\{x\}$.} 
From Equation ({\bf \ref{complete-symmetric}}), 
it follows that
\begin{align}
h_i\{x\} = h_i\{\widehat{x}_m\} + x_m h_{i-1}\{x\}
\label{i1}
\end{align}
From Equation ({\bf \ref{i1}}), one obtains
\begin{align}
(x_m-x_n)
h_{i-1}\{x\}
=
h_i\{\widehat{x}_n\}
-
h_i\{\widehat{x}_m\}
\label{i3}
\end{align}
\begin{align}
(x_m-x_n)
h_i\{x\}
=
x_m
h_i\{\widehat{x}_n\}
-
x_n
h_i\{\widehat{x}_m\}
\label{i2}
\end{align}

\subsection{Discrete derivatives}
The discrete derivative $\Delta_m h_i\{x\}$ of $h_i\{x\}$ 
with respect to $x_m \in \{x\}$ is defined 
using Equation ({\bf \ref{i1}}) as  
\begin{align} 
\Delta_{m} h_i\{x\} = 
\frac{
h_i\{x\} - h_i\{ \widehat{x}_m \}
}
{x_m
} 
= h_{i-1}\{x\}
\label{discrete-derivative}
\end{align}
Note that by applying $\Delta_m$ to a degree $i$ complete symmetric 
function, $h_i\{x\}$, one obtains a complete symmetric function 
$h_{i-1}\{x\}$ of degree $i-1$, in the same set of variables $\{x\}$.

\subsection{The discrete KP hierarchy}

Discrete KP is an infinite hierarchy of integrable 
partial {\it difference} equations in an infinite set of continuous 
Miwa variables $\{x_1,x_2,\dots\}$ with multiplicities $\{m_1,m_2,\dots\}$. 
Time evolution is obtained by changing 
the multiplicities of the Miwa variables. In this work, we take the number 
of non-zero Miwa 
variables to be finite, and set all continuous 
Miwa variables apart from $\{x_1,\dots,x_N\}$ to zero. 
In this case, the discrete KP hierarchy can be written in bilinear form 
as the $(n\! \times \! n)$ determinant equations
\begin{align}
\det
\ll
\begin{array}{cccccc}
     1 & x_1    & \cdots & x_1^{n-2} & \, & x_1^{n-2} \tau_{+1}\{x\}
\tau_{-1}\{x\} \\
     1 & x_2    & \cdots & x_2^{n-2} & \, & x_2^{n-2} \tau_{+2}\{x\}
\tau_{-2}\{x\} \\
\vdots & \vdots & \vdots & \vdots    & \, & \vdots                         
        \\
     1 & x_n    & \cdots & x_n^{n-2} & \, & x_n^{n-2} \tau_{+n}\{x\}
\tau_{-n}\{x\} 
\end{array}
\rr
=0
\label{bilinear-difference-equation}
\end{align}
where $3 \leq n \leq N$, and
\begin{align}
\tau_{+i}\{x\} 
&= 
\tau\{ x_1^{(m_1    )}, \dots, x_i^{(m_i + 1)}, \dots, x_N^{(m_N    )}\}
\\
\tau_{-i}\{x\} 
&= 
\tau\{ x_1^{(m_1 + 1)}, \dots, x_i^{(m_i    )}, \dots, x_N^{(m_N + 1)}\}
\nonumber
\end{align}
In words, if $\tau\{x\}$ has $m_i$ copies of the variable $x_i$, then 
$\tau_{+i}\{x\}$ has $(m_i + 1)$ copies of $x_i$ and the multiplicities 
of all other variables remain the same, while $\tau_{-i}\{x\}$ has one 
more copy of each variable except $x_i$. In the simpler notation

\begin{align}
\tau_{+i}\{x\} 
&= 
\tau\{  m_1,      \dots, (m_i + 1), \dots,  m_N     \}
\label{notation-for-tau}
\\
\tau_{-i}\{x\} 
&=
\tau\{ (m_1 + 1), \dots,  m_i     , \dots, (m_N + 1)\}
\nonumber
\end{align}
the simplest discrete KP bilinear difference equation is 
\begin{multline}
\label{miwa-hirota}
x_i(x_j-x_k) \tau\{ m_i+1, m_j,   m_k   \} \tau\{ m_i,   m_j+1, m_k+1 \} 
\\
+ 
x_j(x_k-x_i) \tau\{ m_i,   m_j+1, m_k   \} \tau\{ m_i+1, m_j,   m_k+1 \}
\\
+ 
x_k(x_i-x_j) \tau\{ m_i,   m_j,   m_k+1 \} \tau\{ m_i+1, m_j+1, m_k   \}   
= 0
\end{multline}
where $\{x_i, x_j, x_k\} \in \{x\}$ and 
$\{m_i, m_j, m_k\} \in \{m\}$ are
any three continuous Miwa variables and their corresponding 
multiplicities.

\subsection{Casoratian matrices and determinants}

$\Omega$ is a Casoratian matrix if and only if its matrix elements 
$\omega_{ij}$ satisfy 
\begin{align}
\omega_{i,j+1}\{x\} = \Delta_m \ \omega_{ij}\{x\} 
\label{casoratian-condition}
\end{align}
where $\Delta_m$ is the discrete derivative with 
respect to any variable $x_m \in \{x\}$. It is redundant to choose a 
specific variable $x_m$, since $\omega_{ij}\{x\}$ is symmetric in $\{x\}$.

From the definition of $\Delta_m$, the elements $\omega_{ij}$ of 
Casoratian matrices satisfy
\begin{align}
\label{a1}
\omega_{ij}\{x_1, \ldots, x_{m}^{(2)}, \ldots, x_{N}\}
=
\omega_{i j  }\{x_1, \ldots,                      x_N \} 
%\\
+ 
x_m \omega_{i,j+1}\{x_1, \ldots, x^{(2)}_{m}, \ldots, x_N \}
\end{align}
which gives the identity
\begin{multline}
(x_r - x_s) 
\  
\omega_{ij} \{x_{1}, \ldots, 
              x^{(2)}_{r}, \dots, x^{(2)}_{s}, \ldots x_N \} = 
\\
x_{r} \ \omega_{ij} \{x_1,   \ldots, x_{r}^{(2)}, \ldots, x_{N} \}
-
x_{s} \ \omega_{ij} \{x_1,   \ldots, x_{s}^{(2)}, \ldots, x_{N} \}
\label{a2}
\end{multline}
If $\Omega$ is a Casoratian matrix, then $\det \Omega$ 
is a Casoratian determinant. Casoratian determinants are 
discrete analogues of Wronskian determinants.

\subsection{Notation for column vectors and determinants}

We introduce the column vector notation
\begin{align}
\vec\omega_{j} =
\ll
\begin{array}{c} 
\omega_{1j} \{ x^{(m_{1})}_1, \ldots, x^{(m_{N})}_{N} \} \\ 
\omega_{2j} \{ x^{(m_{1})}_1, \ldots, x^{(m_{N})}_{N} \} \\ 
\vdots                               \\ 
\omega_{Nj} \{ x^{(m_{1})}_1, \ldots, x^{(m_{N})}_{N} \} 
\end{array}
\rr
\end{align}
and 
\begin{align}
\vec\omega_{j}^{[k_1,\ldots,k_n]} =
\ll
\begin{array}{c} 
\omega_{1j}\{ x^{(m_{1})}_1, \ldots, x^{(m_{k_1}+1)}_{k_1},
                             \ldots, x^{(m_{k_n}+1)}_{k_n},
                             \ldots, x^{(m_{N})}_{N} \} \\ 
\omega_{2j}\{ x^{(m_{1})}_1, \ldots, x^{(m_{k_1}+1)}_{k_1},
                             \ldots, x^{(m_{k_n}+1)}_{k_n},
                             \ldots, x^{(m_{N})}_{N} \} \\ 
\vdots \\ 
\omega_{Nj}\{ x^{(m_{1})}_1, \ldots, x^{(m_{k_1}+1)}_{k_1},
                             \ldots, x^{(m_{k_n}+1)}_{k_n},
                             \ldots, x^{(m_{N})}_{N} \}
\end{array}
\rr
\end{align}
for the corresponding column vector where the multiplicity 
of the subset of variables $x_{k_1}, $$\dots,$ $x_{k_n}$ 
is increased by 1. We introduce the determinant notation 

\begin{align}
\tau= 
\det 
\ll   \vec\omega_{1}\,\, 
      \vec\omega_{2}\,\,
      \cdots        \,\,
      \vec\omega_{N}
\rr
=
\big | \, 
      \vec\omega_{1}\,\, 
      \vec\omega_{2}\,\,
      \cdots        \,\,
      \vec\omega_{N}
\, \big |
\end{align}
and 
\begin{align}
\tau^{[k_1, \ldots, k_n]} 
=
\big | \, 
\vec\omega_{1}^{[k_1, \ldots, k_n]} \, \, 
\vec\omega_{2}^{[k_1, \ldots, k_n]} \, \, 
\cdots                           \, \, 
\vec\omega_{N}^{[k_1, \ldots, k_n]}
\, \big |
\label{defn}
\end{align}
for the determinant with shifted multiplicities. 

\subsection{Identities for Casoratian determinants}

Following \cite{FS}, Equations ({\bf \ref{a1}}) and ({\bf \ref{a2}}) can 
be used to perform column operations in the determinant expressions for 
$\tau^{[1]}$ and $\tau^{[1,\dots,n]}$, to obtain the two identities
 
\begin{align}
x^{n-2}_{1} \ \tau^{[1]} =
\big | \,
\vec\omega_{1}   \,\,
\vec\omega_{2}   \,\,
\cdots           \,\, 
\vec\omega_{N-1} \,\,
\vec\omega_{N-n+2}^{[1]}
\, \big |
\label{A1}
\end{align}
\begin{align}
\prod_{1\leq r<s\leq n}(x_r-x_s)  \tau^{[1, \ldots, n]} =
\big | \,
\vec\omega_1            \,\,
\ldots                  \,\,
\vec\omega_{N-n}        \,\,
\vec\omega_{N-n+1}^{[n]}   \,\,
\vec\omega_{N-n+1}^{[n-1]} \,\,
\ldots                  \,\,
\vec\omega_{N-n+1}^{[1]}
\, \big | 
\label{A2}
\end{align}

\subsection{Casoratian determinants are discrete KP $\tau$-functions}
Following \cite{ohta}, consider the $(2N\! \times \! 2N)$ determinant

\begin{align}
\det
\ll
\begin{array}{cccccccc}
\vec\omega_1          & 
\cdots                &
\vec\omega_{N-1}      & 
\vec\omega_{N-n+2}^{[1]} &    
0_1                   & 
\cdots                &    
0_{N-n+1}             &
\vec\omega_{N-n+2}^{[n]} 
\cdots 
\vec\omega_{N-n+2}^{[2]} 
\\
0_1                   & 
\cdots                &    
0_{N-1}               & 
\vec\omega_{N-n+2}^{[1]} & 
\vec\omega_1          & 
\cdots                & 
\vec\omega_{N-n+1}    &
\vec\omega_{N-n+2}^{[n]} 
\cdots 
\vec\omega_{N-n+2}^{[2]} 
\end{array}
\rr
= 0
\label{2N-by-2N}
\end{align}
which is identically zero. For notational clarity, we have used subscripts 
to label the position of columns of zeros. Laplace expanding the left 
hand side of Equation ({\bf \ref{2N-by-2N}}) in 
$(N\! \times \! N)$ minors along the top 
$(N\! \times \! 2N)$ block, we obtain 
\begin{multline}
\sum_{k = 1}^{n} 
(-)^{k - 1}
\big | \,
  \vec\omega_1 
  \cdots 
  \vec\omega_{N-1}  
  \vec\omega_{N-n+2}^{[k]} 
\, \big | 
\times \\
\big | \,
  \vec\omega_1 
  \cdots 
  \vec\omega_{N-n+1}         
  \vec\omega_{N-n+2}^{[n]} 
  \cdots 
  \vec\omega_{N-n+2}^{[k+1]} 
  \vec\omega_{N-n+2}^{[k-1]} 
  \cdots 
  \vec\omega_{N-n+2}^{[1]}
\, \big |
= 0
\label{Laplace-expansion}
\end{multline}
From Equations ({\bf \ref{A1}}) and ({\bf \ref{A2}}), 
Equation ({\bf \ref{Laplace-expansion}}) can be written as
\begin{align}
\sum_{k = 1}^{n} 
(-)^{k - 1}
x^{n-2}_{k}\ \tau^{[k]}
\prod_{\substack{1\leq r<s\leq n \\ r,s\neq k}}(x_r-x_s)
\tau^{[1,\ldots\hat k \ldots,n]}
= 0
\label{57}
\end{align}
Using the Vandermonde determinant identity
\begin{align}
\det
\ll
\begin{array}{cccccc}
     & 1 & x_1    & \cdots & x_1^{n-2} & \\
     & \vdots & \vdots & & \vdots & \\
\langle  & 1 & x_{k}  & \cdots & x_{k}^{n-2} & \rangle \\
	& \vdots & \vdots & & \vdots &  \\
      & 1 & x_n    & \cdots & x_n^{n-2}  &
\end{array}
\rr
=
\prod_{\substack{1\leq r<s\leq n \\ r,s\neq k}}(x_r-x_s)
\end{align}
with 
$\langle 
\begin{array}{cccc} 1 & x_k & \cdots & x_k^{n-2} \end{array} 
\rangle$ 
denoting the omission of the $k$-th row of the matrix, we see that 
Equation ({\bf \ref{57}}) is the cofactor expansion of the determinant in 
Equation ({\bf \ref{bilinear-difference-equation}}) along its last 
column. Hence we conclude that Casoratian determinants satisfy 
the bilinear difference equations of discrete KP.

\subsection{$\zi ( \{x\}_n | \{y\}_N )$ is a discrete KP $\tau$-function 
in $\{y\}_N$}

In this subsection we show that $\zi ( \{x\}_n | \{y\}_N )$, in Equation 
({\bf \ref{poly1}}), is a Casoratian determinant. The discrete derivatives 
are taken with respect to any of the variables $y_j$. From the above 
discussion, this is sufficient to show that $\zi ( \{x\}_n | \{y\}_N )$ is 
a $\tau$-function of discrete KP in $\{y\}_N$.

The first step is to rearrange Equation ({\bf \ref{poly1}}) by bringing 
the numerator of the prefactor $\mathcal{P}(\{x\}_n,\{y\}_N)$ inside the 
determinant. We do this by multiplying the $j$-th column of the 
determinant by $\prod_{k=1}^{n} (x_k-y_j)(x_k-\bary_j)$, for all 
$1 \leq j \leq N$.  The $j$-th column of the resulting determinant 
has entries which are polynomial in $y_j$. After a routine calculation, 
we obtain
\begin{multline}
\zi \ll \{x\}_n \Big| \{y\}_N \rr
=
\Delta^{-1}\{ x\}_n \ 
\Delta^{-1}\{-y\}_N \ 
\det
\ll
\sum_{k=1}^{N+n}
c_{ik}\{x\} y_j^{k-1}
\rr_{1 \leq i,j \leq N}
\label{IK-tau1}
\end{multline}
where the coefficients $c_{ik}\{x\}$ depend on the row of the 
matrix and are given by 
\begin{align}
c_{ik}\{x\}
=
\left\{
\begin{array}{ll}
e_{2n-k-1}
\ll
\{-x,-\barx\}
\backslash
\{-x_i,-\barx_i\}
\rr,
&
\quad
1 \leq i \leq n
\\
\\
e_{2n-k+N-i+1}
\{-x,-\barx\},
&
\quad
n+1 \leq i \leq N
\end{array}
\right.
\end{align}
It remains to take the Vandermonde $\Delta\{-y\}_N$ inside 
the determinant of Equation ({\bf \ref{IK-tau1}}). This is 
essentially the same as proving the Jacobi-Trudi identity 
for Schur functions, see \cite{macdonald.book}. The final 
result is

\begin{equation}
\zi \ll \{x\}_n \Big| \{y\}_N \rr
=
\Delta^{-1}\{x\}_n
\ 
\det
\ll 
\sum_{k=1}^{N+n} c_{ik}\{x\} h_{k-j}\{y\} 
\rr_{1\leq i,j \leq N}
\label{tau-ik}
\end{equation}
Up to the Vandermonde factor in the denominator, which 
is a constant in $\{y\}_N$, this is clearly a Casoratian 
determinant.

\subsection{$\zs ( \{x\}_n | \{y\}_N )$ is a discrete KP 
$\tau$-function in $\{x\}_n$}

We can repeat the above procedure to write $\zs ( \{x\}_n | \{y\}_N )$ 
as a Casoratian determinant, whose discrete derivatives are with respect 
to any of the variables $x_i$. Starting from Equation ({\bf \ref{poly2}}), 
we already have $\zs ( \{x\}_n | \{y\}_N )$ as a determinant whose $i$-th 
row entries are polynomials in $x_i$. Expanding these polynomials in powers 
of $x_i$, we obtain
\begin{align}
\zs \ll \{x\}_n \Big| \{y\}_N \rr
=
\Delta^{-1}\{x\}_n \ 
\det
\ll
\sum_{k=1}^{N+n}
x_i^{k-1}
d_{kj}\{y\}
\rr_{1\leq i,j \leq n}
\label{S-tau1}
\end{align}
where the coefficients $d_{kj}\{y\}$ depend on the column of 
the matrix and are given by

\begin{align}
d_{kj}\{y\}
=
\sum_{l=0}^{N+j-k}
\left[
\binom{N-l}{k-j}
-
\binom{j-1}{k-N+l-1}
\right]
e_l\{-y\}
\end{align}
Taking the Vandermonde $\Delta\{x\}_n$ inside the determinant of 
Equation ({\bf \ref{S-tau1}}), we have

\begin{align}
\zs \ll \{x\}_n \Big| \{y\}_N \rr
=
\det
\ll
\sum_{k=1}^{N+n}
h_{k-i}\{x\}
d_{kj}\{y\}
\rr_{1 \leq i,j \leq n}
\end{align}
Hence $\zs ( \{x\}_n | \{y\}_N )$ is a Casoratian determinant, 
and satisfies the discrete KP equations in $\{x\}_n$.

\section{The Gromov-Vieira polynomial version of partial 
domain wall partition functions}
\label{section-6}

Following \cite{GSV}, partial domain wall configurations are 
(the essential part of) 3-point functions of tree-level 
single-trace operators in the $SU(2)$ sector of ${\rm SYM}_4$,
with two BPS and one non-BPS operators. 
In \cite{gromov.vieira.short, gromov.vieira.long}, Gromov and 
Vieira showed that 1-loop corrections can be introduced using 
the mapping discussed in this section. In the sequel, we show 
that the determinant form of these objects at tree-level is 
preserved under the GV mapping, thus the corresponding 1-loop 
corrected objects in SYM$_4$ can also be expressed 
as determinants. 

\subsection{The Gromov-Vieira mapping}

In \cite{gromov.vieira.short, gromov.vieira.long}, Gromov and 
Vieira define the following mapping on any function 
$f(\theta_1,\dots,\theta_N)$ of the variables 
$\{\theta_1,\dots,\theta_N\}$,

\begin{align}
f \mapsto
[ f ]_{\theta}
=
f
\Big|_{\theta_1,\dots,\theta_N \rightarrow 0}
+
\frac{g^2}{2}
\sum_{i=1}^{N}
(\partial_{\theta_i} - \partial_{\theta_{i+1}})^2
f 
\ 
\Big|_{\theta_1,\dots,\theta_N \rightarrow 0}
+
O(g^4)
\label{theta}
\end{align}
where $\partial_{\theta_{N+1}} \equiv \partial_{\theta_1}$. 
Note that the mapping is defined to $O(g^2)$ in some small 
expansion parameter $g$. 

\subsection{Aim of this section}

Our aim is to show that, up to $O(g^2)$, the GV mapping acts on 
a Casoratian determinant $\zi ( \{x\}_n | \{y\}_N )$ to return 
a new determinant. We show this by explicitly evaluating
$
\left[
\zi ( \{x\}_n | \{y\}_N )
\right]_y
$.

\subsection{The GV mapping in terms of symmetric functions}
We need the following degree-2 cyclically symmetric function
in $\{y\} = \{y_1, y_2, \dots, y_N\}$,

\begin{equation}
m_2 \{y\} = y_1 y_2 + y_2 y_3 + \cdots + y_{(N-1)} y_N + y_N y_1
\label{m-function}
\end{equation}

\noindent Using $m_2 \{y\}$ and the definition of the complete 
symmetric functions in Equation ({\bf \ref{complete-symmetric}}), 
one can write 

\begin{equation}
\label{identity}
\sum_{i=1}^{N}
(y_i - y_{i+1})^2
=
4 h_2\{y\} - 2 h_1^2\{y\} - 2 m_2\{y\}
\end{equation}
where we assume the periodicity $y_{(N+1)} \equiv y_1$. In terms
of the corresponding differential operators, 

\begin{multline}
\sum_{i=1}^{N}
(\partial_{y_i} - \partial_{y_{i+1}})^2
=
4 h_2\{\partial_y\} - 2 h_1^2\{\partial_y\} 
                    - 2     m_2\{\partial_y\}
\\
=
4 h_2\{\partial_y\} - 2 \ll h_1^2\{\partial_y\} + m_2\{\partial_y\} \rr
=
4 h_2\{\partial_y\} - 2 g_2\{\partial_y\} 
\end{multline}

\noindent where we have defined 

\begin{equation}
g_2\{\partial_y\} = h_1^2\{\partial_y\} + m_2\{\partial_y\}
\end{equation}

\noindent We are interested in computing 
$ h_2\{\partial_y\}    f\{y\} |_{y_1,\dots,y_N \rightarrow 0}$, and
$ g_2\{\partial_y\}    f\{y\} |_{y_1,\dots,y_N \rightarrow 0}$, 
for generic symmetric functions $f\{y\}$, so for convenience we 
adopt the shorthand
\begin{align}
    h_2\{\partial_y\}       f\{y\} |_{y_1, \dots, y_N \rightarrow 0} 
\equiv H_2 f,
\quad
    g_2\{\partial_y\}       f\{y\} |_{y_1, \dots, y_N \rightarrow 0} 
\equiv G_2 f,
\label{H-op}
\end{align}

\subsection{Action of $H_2$, and $G_2$}
\label{action}
 
Let $(h_{1})^{m_1} (h_{2})^{m_2} \dots (h_{L})^{m_L}$ be an arbitrary 
monomial in the complete symmetric functions. Using the definitions 
in Equation ({\bf \ref{H-op}}), we obtain

\begin{multline}
\label{H2}
H_2 \ll (h_{1})^{m_1} (h_{2})^{m_2} \dots (h_{L})^{m_L} \rr
=
\\
\left\{
\begin{array}{ll}
N(N+1), & m_1 = 2, m_2 = 0, m_3 = \cdots = m_L = 0 
\\
N(N+3)/2, & m_1 = 0, m_2 = 1, m_3 = \cdots = m_L = 0
\\
0, & {\rm otherwise}
\end{array}
\right.
\end{multline}

\begin{multline}
\label{G2}
G_2 \ll (h_{1})^{m_1} (h_{2})^{m_2} \dots (h_{L})^{m_L} \rr
=
\\
\left\{
\begin{array}{ll}
2N(N+1), & m_1 = 2, m_2 = 0, m_3 = \cdots = m_L = 0 
\\
N(N+2), & m_1 = 0, m_2 = 1, m_3 = \cdots = m_L = 0
\\
0, & {\rm otherwise}
\end{array}
\right.
\end{multline}

\noindent Both $H_2$ and $G_2$ act trivially on any monomial whose 
degree $d = m_1 + 2 m_2 + \cdots + L m_L \not= 2$, which greatly 
simplifies the action of the GV mapping on $\zi ( \{x\}_n | \{y\}_N )$.

\subsection{Remarks on notation}

Henceforth we reserve $i$ and $j$ for the row and column indices 
of a determinant, respectively, and assume that they range over all 
values $1 \leq i,j \leq N$. For example, we write the determinant 
in Equation ({\bf \ref{tau-ik}}) as  

\begin{align}
\left|
\sum_{k=1}^{N+n} c_{ik}\{x\} h_{k-j}\{y\}
\right|
=
\left|
\sum_{k=1}^{N+n} c_{jk}\{x\} h_{k-i}\{y\}
\right|
=
\left|
\begin{array}{c}
c_{jk} h_{k-1}
\\
\vdots
\\
c_{jk} h_{k-N}
\end{array}
\right|
\end{align}
where the first equality follows from the invariance of the determinant 
under matrix transposition. In the second equality we suppress arguments 
and the summation symbol, but show the $j$-th row explicitly.  In the 
rest of this section, all calculations will change determinant on 
a row-by-row basis.

\subsection{Degree-2 terms in the determinant}

Since the only terms which survive under the action of $H_2$ and $G_2$ 
are degree-2 monomials in the complete symmetric functions, we focus 
on these terms by expanding our determinant as follows

\begin{align}
\left|
\begin{array}{c}
c_{j,k} h_{k-1}
\\
\vdots
\\
c_{j,k} h_{k-N}
\end{array}
\right|
=
\sum_{l=1}^{N}
\left|
\begin{array}{c}
c_{j,1} h_0
\\
\vdots
\\
c_{j,l-1} h_0
\\
c_{j,l+2} h_2
\\
c_{j,l+1} h_0
\\
\vdots
\\
c_{j,N} h_0
\end{array}
\right|
+
\sum_{1 \leq l_1 < l_2 \leq N}
\left|
\begin{array}{c}
c_{j,1} h_0
\\
\vdots
\\
c_{j,l_1-1} h_0
\\
c_{j,l_1+1} h_1
%\\
%c_{j,l_1+1} h_0
\\
\vdots
%\\
%c_{j,l_2-1} h_0
\\
c_{j,l_2+1} h_1
\\
c_{j,l_2+1} h_0
\\
\vdots
\\
c_{j,N} h_0
\end{array}
\right|
+
\text{degree-$(d\not=2)$ terms}
\label{deg-2}
\end{align}
where we maintain the symbol $h_0$ for clarity, despite the 
fact that $h_0 =1$. 

\subsection{Action of $H_2$ on Equation ({\bf \ref{deg-2}})}

Acting on Equation ({\bf \ref{deg-2}}) with $H_2$ and using 
Equation ({\bf \ref{H2}}), we find 

\begin{multline}
H_2
\left|
\begin{array}{c}
c_{j,k} h_{k-1}
\\
\vdots
\\
c_{j,k} h_{k-N}
\end{array}
\right|
=
\frac{N(N+3)}{2}
\left|
\begin{array}{c}
c_{j,1}
\\
\vdots
\\
c_{j,N-2}
\\
c_{j,N+1}
\\
c_{j,N}
\end{array}
\right|
+
\frac{N(N+3)}{2}
\left|
\begin{array}{c}
c_{j,1}
\\
\vdots
\\
c_{j,N-1}
\\
c_{j,N+2}
\end{array}
\right|
+
N(N+1)
\left|
\begin{array}{c}
c_{j,1}
\\
\vdots
\\
c_{j,N-2}
\\
c_{j,N}
\\
c_{j,N+1}
\end{array}
\right|
\label{action-1}
\end{multline}
The first two terms come from the sum $\sum_{l=1}^{N}$ in 
Equation ({\bf \ref{deg-2}}), while the final term comes 
from the sum $\sum_{1 \leq l_1 < l_2 \leq N}$. All other 
terms vanish under the action of $H_2$, either because 
they have the wrong degree or give rise to a determinant 
with two equivalent rows.

Combining the first and third determinant in 
Equation ({\bf \ref{action-1}}), which are the same up 
to the ordering of their rows, we obtain

\begin{align}
H_2
\left|
\begin{array}{c}
c_{j,k} h_{k-1}
\\
\vdots
\\
c_{j,k} h_{k-N}
\end{array}
\right|
=
-
\frac{N(N-1)}{2}
\left|
\begin{array}{c}
c_{j,1}
\\
\vdots
\\
c_{j,N-2}
\\
c_{j,N+1}
\\
c_{j,N}
\end{array}
\right|
+
\frac{N(N+3)}{2}
\left|
\begin{array}{c}
c_{j,1}
\\
\vdots
\\
c_{j,N-1}
\\
c_{j,N+2}
\end{array}
\right|
\label{action-3}
\end{align} 

\subsection{Action of $G_2$ on Equation ({\bf \ref{deg-2}})}

Acting on Equation ({\bf \ref{deg-2}}) with $G_2$ and using 
Equation ({\bf \ref{G2}}), we find 
\begin{align}
G_2
\left|
\begin{array}{c}
c_{j,k} h_{k-1}
\\
\vdots
\\
c_{j,k} h_{k-N}
\end{array}
\right|
&=
N(N + 2)
\left|
\begin{array}{c}
c_{j,1}
\\
\vdots
\\
c_{j,N-2}
\\
c_{j,N+1}
\\
c_{j,N}
\end{array}
\right|
+
N(N + 2)
\left|
\begin{array}{c}
c_{j,1}
\\
\vdots
\\
c_{j,N-1}
\\
c_{j,N+2}
\end{array}
\right|
+
2N(N + 1)
\left|
\begin{array}{c}
c_{j,1}
\\
\vdots
\\
c_{j,N-2}
\\
c_{j,N}
\\
c_{j,N+1}
\end{array}
\right|
\label{action-2}
\end{align}
The first two terms come from the sum $\sum_{l=1}^{N}$ in 
Equation ({\bf \ref{deg-2}}), while the final term comes 
from the sum $\sum_{1 \leq l_1 < l_2 \leq N}$. All other 
terms vanish under the action of $G_2$, either because 
they have the wrong degree or give rise to a determinant 
with two equivalent rows.

Combining the first and third determinant in 
Equation ({\bf \ref{action-2}}), which are the same up 
to the ordering of their rows, we obtain

\begin{align}
G_2
\left|
\begin{array}{c}
c_{j,k} h_{k-1}
\\
\vdots
\\
c_{j,k} h_{k-N}
\end{array}
\right|
&=
-
N^2
\left|
\begin{array}{c}
c_{j,1}
\\
\vdots
\\
c_{j,N-2}
\\
c_{j,N+1}
\\
c_{j,N}
\end{array}
\right|
+
N(N + 2)
\left|
\begin{array}{c}
c_{j,1}
\\
\vdots
\\
c_{j,N-1}
\\
c_{j,N+2}
\end{array}
\right|
\label{action-4}
\end{align}

\subsection{Determinant expression for 
$[\zi ( \{x\}_n | \{y\}_N ) ]_y$}

We are ready to express the action of the GV mapping on 
$\zi ( \{x\}_n | \{y\}_N )$ as a single determinant. 
Firstly, using Equation ({\bf \ref{tau-ik}}) it is trivial 
to calculate
\begin{align}
\zi \ll \{x\}_n \Big| \{y\}_N \rr
\Big|_{y_1,\dots,y_N \rightarrow 0}
=
\Delta^{-1} \{x\}_n 
\ 
\left|
\begin{array}{c}
c_{j,1} 
\\
\vdots
\\
c_{j,N} 
\end{array}
\right|
\label{1}
\end{align}
where the column index $j$ ranges over all values $1 \leq j \leq N$, 
as usual. For the second part of the GV mapping in 
Equation ({\bf \ref{theta}}), we wish to calculate
\begin{align}
\sum_{i=1}^{N}
(\partial_{y_i} - \partial_{y_{i+1}})^2
\zi \ll \{x\}_n \Big| \{y\}_N \rr
\Big|_{y_1,\dots,y_N \rightarrow 0}
=
\frac{
\left(
4 H_2 - 2 G_2
\right)
}{
\Delta\{x\}_n
}
\left|
\begin{array}{c}
c_{j,k} h_{k-1}
\\
\vdots
\\
c_{j,k} h_{k-N}
\end{array}
\right|
\label{2}
\end{align}
Putting together the results of the previous subsections, 
namely Equations ({\bf \ref{action-3}}) and 
({\bf \ref{action-4}}), we find that
\begin{align}
\frac{
\left(
4 H_2 - 2 G_2
\right)
}{
\Delta\{x\}_n
}
\left|
\begin{array}{c}
c_{j,k} h_{k-1}
\\
\vdots
\\
c_{j,k} h_{k-N}
\end{array}
\right|
=
\frac{2N}
{
\Delta\{x\}_n
}
\left|
\begin{array}{c}
c_{j,1}
\\
\vdots
\\
c_{j,N-2}
\\
c_{j,N+1}
\\
c_{j,N}
\end{array}
\right|
+
\frac{2N}
{
\Delta\{x\}_n
}
\left|
\begin{array}{c}
c_{j,1}
\\
\vdots
\\
c_{j,N-1}
\\
c_{j,N+2}
\end{array}
\right|
\label{3}
\end{align}
Using Equations ({\bf \ref{1}}--{\bf \ref{3}}) we obtain 

\begin{multline}
\label{4}
\left[ \zi \ll \{x\}_n \Big| \{y\}_N \rr \right]_y
\equiv
\\
\zi
\Big|_{y_1,\dots,y_N \rightarrow 0}
+
\frac{g^2}{2}
\sum_{i=1}^{N}
(\partial_{y_i} - \partial_{y_{i+1}})^2
\zi
\Big|_{y_1,\dots,y_N \rightarrow 0}
+
O(g^4)
\\
=
\frac{1}{\Delta\{x\}_n}
\left|
\begin{array}{c}
c_{j,1}
\\
\vdots
\\
c_{j,N}
\end{array}
\right|
+
\frac{g^2 N}{\Delta\{x\}_n}
\left|
\begin{array}{c}
c_{j,1}
\\
\vdots
\\
c_{j,N-2}
\\
c_{j,N+1}
\\
c_{j,N}
\end{array}
\right|
+
\frac{g^2 N}{\Delta\{x\}_n}
\left|
\begin{array}{c}
c_{j,1}
\\
\vdots
\\
c_{j,N-1}
\\
c_{j,N+2}
\end{array}
\right|
+
O(g^4)
\end{multline}
The first three terms of Equation ({\bf \ref{4}}) can actually 
be combined into a single determinant, which is correct up to 
$O(g^2)$. Our final result is

\begin{align}
\left[ \zi \ll \{x\}_n \Big| \{y\}_N \rr \right]_y
=
\frac{1}{\Delta\{x\}_n}
\left|
\begin{array}{c}
c_{j,1}
\\
\vdots
\\
c_{j,N-2}
\\
c_{j, N-1} + g^2 N c_{j,N+1}
\\
c_{j, N\phantom{-1}}   + g^2 N c_{j,N+2}
\end{array}
\right|
+
O(g^4)
\label{final-exp-gv}
\end{align}

The result in Equation ({\bf \ref{final-exp-gv}}) is such a simple 
modification of the original expression, obtained by setting 
$g^2 \to 0$, that we expect that higher derivative versions of the 
GV mapping will also preserve the determinant form of the pDWPF
\footnote{\  An earlier draft of this work contained an incorrect 
version of Equation ({\bf {\ref{identity}}}) that led to a more 
complicated version of Equation ({\bf \ref{final-exp-gv}}).
We thank D Serban for pointing this out.}. 

Since the GV mapping is an expansion around 
the homogeneous limit at which all variables $y_i =0$, we cannot 
consider the determinant in Equation ({\bf\ref{final-exp-gv}}) to 
be a discrete KP $\tau$-function in the $\{y\}$ variables. On the 
other hand, according to the methods of Section {\bf 5}, the 
determinant in Equation ({\bf \ref{final-exp-gv}}) is not in 
Casorati form, hence we cannot conclude that it is a discrete 
KP $\tau$-function in the $\{x\}$ variables.

\section{Remarks}
\label{section-7}

\subsection{Summary of results}
Rational and trigonometric partial domain wall partition 
functions, pDWPF's, are partition functions of six-vertex 
model configurations on lattices with unequal numbers of 
horizontal lines $L_h$ and vertical lines $L_v$. 
They can be regarded as less restrictive variations on 
Korepin's rational and trigonometric domain wall partition 
functions, DWPF's, which require $L_h = L_v$, but can be 
deduced from them, as well as from configurations that
describe scalar products, by taking some of the rapidities 
to infinity. 

In this work, we gave explicit derivations of the determinant 
expressions for pDWPF's as limits of Izergin's DWPF determinant, 
as well as of Slavnov's determinant for the scalar product of 
a Bethe eigenstate and a generic state, in the rational and 
trigonometric cases, and studied some of their properties. 
The rational pDWPF was first derived from Slavnov's determinant 
by I Kostov \cite{kostov.private.communication}.
We showed how the two determinants obtained as limits of Izergin's 
determinant and of Slavnov's determinant are different (one is 
$(N\! \times \! N)$ while the other is $(n\! \times \! n)$, where 
$n < N$), but 
can be directly related, that they are KP $\tau$-functions in 
each of two sets of variables, and that they remain determinants 
under the mapping of Gromov and Vieira 
\footnote{\  It is likely that the determinant expression is 
preserved under the action of higher derivative versions of 
the GV mapping. We did not pursue this since, at this stage, 
the relation between the higher derivative versions and 
the inclusion of higher loop corrections to the 3-point 
functions is not clear. However, in \cite{serban}, D Serban 
argued that this is indeed the case, at least in the limit 
$L_i \to \infty$, $i \in \{1, 2, 3\}$. That is, when all 
three operators are represented by asymptotically long 
spin chain states.}.

\subsection{Taking the free variables to infinity in Slavnov's 
determinant}
In Section {\bf 3}, following Kostov
\cite{kostov.private.communication}, we derived pDWPF's from 
Slavnov's scalar products. We kept the free rapidity variables 
$\{x\}$ finite, and took the rapidity variables that satisfy 
Bethe equations, $\{b\}$, to infinity. 
The result is finite and non-trivial.

If we would have kept the Bethe roots $\{b\}$ finite and took 
$\{x\}$ to infinity, the result would have been zero. 
The reason is that this limit corresponds to the scalar product 
of a Bethe eigenstate, labeled by $\{b\}$, and a descendant of 
the reference state (the result of the action of spin-lowering 
operators on the reference state, that lower the net spin but 
do not introduce Bethe roots \cite{E1}). Since the scalar 
product of the Bethe eigenstate $| \{b\} \rangle$ and the 
reference state vanishes, the scalar product of $| \{b\} \rangle$
with a descendant of the reference state also vanishes. 
In other words, a pDWPF with auxiliary space (horizontal 
line) rapidities that obey Bethe equations vanishes.

\subsection{Asymptotics}
In \cite{GSV, kostov.private.communication, kostov.short.paper, 
kostov.long.paper}, pDWPF's were studied in the thermodynamic 
limit $L_v \to \infty$, such that the ratios $L_h / L_v$ and 
$x_i / L_v$, $i \in \{1, \dots, L_h\}$, remain finite, where 
$L_v$ ($L_h$) is the number of vertical (horizontal) lattice 
lines,  $L_h < L_v$, and $\{x\}$ are the rapidities of the 
horizontal lines
\footnote{\  Because of the condition that $x_i / L_v$,
$i \in \{1, \dots, L_h\}$, remains finite, this limit is also 
known as the {\it \lq Sutherland limit\rq} 
\cite{sutherland, dhar.shastry}.}. 
While, strictly speaking, the variables $\{x\}$ are free,  
in applications, such as computations of 3-point functions
of two BPS and one non-BPS operators in the scalar sector 
of SYM$_4$, they are restricted to obey the Bethe equations
of a spin chain of length $L$, such that $L \neq L_v$. 
For that reason, Bethe Ansatz asymptotics apply, but 
the pDWPF is nonetheless non-vanishing. This is the 
set-up used in 
\cite{GSV, kostov.short.paper, kostov.long.paper}.

Following \cite{GSV, kostov.short.paper}, in the above thermodynamic 
limit, the variables $\{x\}$, which are solutions of Bethe equations 
of a spin chain of length $L > L_v$, $L \sim L_v$, condense on a set 
of contours $\Gamma = \union_k \Gamma_k$, with linear density 
$\rho\{x\}$, $\rho \sim \cO(1)$, $x_i \sim \cO(L_v)$. In the 
homogeneous limit, $y_i = 0, i \in \{1, \dots, L_v\}$, the 
asymptotic pDWPF can be expressed as an exponential of a contour 
integral over a dilogarithm function 

\begin{equation}
\label{asymptotic.pdwpf}
\exp \ll \oint_{\cC} \frac{d z}{2 \pi}
{\rm Li}_2 \ll e^{i q(z)} \rr \rr,
\quad
{\rm Li}_2 (z) = \sum_{n=1}^{\infty} \frac{z^n}{n^2}
\end{equation}

\noindent where $\cC$ encircles $\Gamma$ counter-clockwise, 
and  

\begin{equation}
q(z) = - i \log \ll f(z) \rr + \int_{\Gamma} dy \frac{\rho(y)}{z - y},
\quad 
f(z) = \ll \frac{z - i/2}{z + i/2} \rr^{L}
\end{equation}

\noindent The point we wish to mention here is that in the same 
limit, the Slavnov scalar product factorizes into a product 
of terms that are either the asymptotic pDWPF in Equation 
({\bf \ref{asymptotic.pdwpf}}), or simple variations of it 
\cite{kostov.short.paper, kostov.long.paper}. 
Thus, at least asymptotically, pDWPF's are building blocks 
of scalar products.

\subsection{Higher rank scalar products}
In \cite{caetano.vieira, wheeler.su(3)}, pDWPF's appear as factors in 
certain degenerations of the $SU(3)$-analogue of Slavnov's scalar 
product. One starts from sum expressions for 
the $SU(3)$-analogue of Slavnov's scalar product, takes both 
sets of Bethe roots to infinity \cite{caetano.vieira}, or either one 
(there are two sets of Bethe roots in $SU(3)$-invariant spin 
chains) \cite{wheeler.su(3)}, only to find that the sum expression 
factorizes into determinants that inevitably include one 
or more pDWPF. This factorization, and the appearance of 
pDWPF's as factors, is expected on general grounds to remain 
the case for $SU(N)$-analogues, $N \geq 4$, of Slavnov's scalar 
product. 
Since no determinant expression is known for the $SU(N)$-analogues 
of Slavnov's scalar product, we hope that a deeper understanding 
of the properties of building blocks, such as pDWPF's, will help 
solve this problem.

\subsection{Combinatorics and counting}
Six-vertex model configurations with domain wall boundary 
conditions are in one-to-one correspondence with alternating sign 
matrices, ASM's \cite{kuperberg}. 
Using this observation, Kuperberg counted $(N\! \times \! N)$ ASM's by 
evaluating Izergin's determinant at the combinatoric value of the 
crossing parameter $\gamma = 2\pi i/3$ \cite{kuperberg}.

This leads one to expect that similar arguments can be applied 
to the pDWPF's to count more general objects than ASM's. 
This is not the case, or at least not in an obvious way, 
because the trigonometric weights that we needed to derive 
determinant expressions for the trigonometric pDWPF's, 
Equation ({\bf\ref{trig-wt}}), contain phases that vary 
from configuration to configuration and thereby rule out 
any (straightforward) 1-counting as in the DWPF case. 

\section*{Acknowledgments}
OF wishes to thank J Caetano, N Gromov, I Kostov, D Serban 
and P Vieira for discussions, H Saleur and V Schomerus for 
hospitality at the Institut Henri Poincare, Paris, where 
this work started, and I Kostov and D Serban for remarks 
that helped us improve the manuscript. 
We thank the Australian Research Council for financial 
support.


\begin{thebibliography}{99}

\bibitem{bena.polchinski.roiban}
I Bena, J Polchinski and R Roiban,
\textit{Hidden symmetries of the AdS(5) $\times$ S5 superstring}
Phys Rev {\bf D69} (2004) 046002 (2004)
{\tt arXiv:hep-th/0305116}

\bibitem{minahan.zarembo}
J A Minahan and K Zarembo,
\textit{The Bethe-Ansatz for N = 4 super Yang-Mills},
JHEP {\bf 0303} (2003) 013,
{\tt hep-th/0212208}

\bibitem{beisert.review}
N Beisert {\it et al.},
\textit{Review of AdS/CFT Integrability: An Overview},
{\tt arxiv:1012.3982}, and the reviews that it introduces.

\bibitem{korepin.paper}
V E Korepin,
\textit{Calculation of norms of Bethe wave functions,}
Commun. Math. Phys. {\bf 86} (1982),
391--418

\bibitem{izergin.paper}
A G Izergin,
\textit{Partition function of the six-vertex model in a finite volume,}
Sov Phys Dokl {\bf 32} (1987),
878--879

\bibitem{korepin.book}
V E Korepin, N M Bogoliubov, A G Izergin,
\textit{Quantum inverse scattering method and correlation functions},
Cambridge University Press (1993)

\bibitem{bogoliubov.pronko.zvonarev}
N M Bogoliubov, A G Pronko and M B Zvonarev, 
{\it Boundary correlation functions of the six-vertex model}, 
J Phys A {\bf 35} (2002) 5525--5541,
{\tt arXiv:math-ph/0203025} 

\bibitem{foda.preston}
O Foda and I Preston, 
{\it On the correlation functions of the domain wall six vertex model}, 
J Stat Mech {\bf 0411} (2004) P11001,
{\tt arXiv:math-ph/0409067} 

\bibitem{colomo.pronko.1}
F Colomo and A Pronko, 
{\it On the partition function of the six-vertex model with domain 
wall boundary conditions}, 
J Phys {\bf A37} (2004) 1987--2002, 
{\tt arXiv:math-ph/0309064} 

\bibitem{colomo.pronko.2}
F Colomo and A G Pronko, 
{\it On two-point boundary correlations in the six-vertex model with DWBC}, 
J Stat Mech {\bf 0505} (2005) P05010 
{\tt arXiv:math-ph/0503049} 

\bibitem{colomo.pronko.3}
F Colomo and A G Pronko, 
{\it On the problem of calculation of correlation functions in 
the six-vertex model with domain wall boundary conditions}, 
{\tt arXiv:1111.4353} 

\bibitem{E1}
J Escobedo, N Gromov, A Sever and P Vieira,
\textit{Tailoring Three-Point Functions and Integrability},
JHEP {\bf 2011} Number 9, 28
{\tt arXiv:1012.2475}

\bibitem{E2}
J Escobedo, N Gromov, A Sever and P Vieira,
\textit{Tailoring Three-Point Functions and Integrability II.
Weak/strong coupling match},
JHEP {\bf 2011}, Number 9, 29
{\tt arXiv:1104.5501}

\bibitem{GSV}
N Gromov, A Sever and P Vieira,
\textit{Tailoring Three-Point Functions and Integrability III.
Classical Tunneling},
to appear in JHEP 
{\tt arXiv:1111.2349}

\bibitem{F}
O Foda,
\textit{$\cN =4$ SYM structure constants as determinants},
JHEP {\bf 2012}, Number 3, 96 
{\tt arXiv:1111.4663}

\bibitem{wheeler}
M Wheeler,
\textit{An Izergin--Korepin procedure for calculating scalar products
in six-vertex models},
Nucl Phys {\bf B852} (2011) 468-507
{\tt arXiv:1104.2113}

\bibitem{kostov.private.communication}
I Kostov,
Private communication.

\bibitem{kostov.short.paper}
I Kostov,
\textit{Classical Limit of the Three-Point Function from Integrability},
{\tt arXiv:1203.6180} 

\bibitem{kostov.long.paper}
I Kostov,
\textit{Three-point function of semiclassical states at weak
coupling},
{\tt arXiv:1205.4412}

\bibitem{gromov.vieira.short}
N Gromov and P Vieira,
{\it Quantum integrability for three-point functions,}
{\tt arXiv:1202.4103}

\bibitem{gromov.vieira.long}
N Gromov and P Vieira,
{\it Tailoring Three-Point Functions and Integrability IV.
Theta--Morphism}, 
{\tt  arXiv:1205.5288}

\bibitem{slavnov}
N A Slavnov,
\textit{Calculation of scalar products of wave functions and 
form factors in the framework of the algebraic Bethe Ansatz,}
Theor Math Phys {\bf 79} (1989),
502--508

\bibitem{KMT}
N Kitanine, J M Maillet and V Terras,
\textit{Form factors of the XXZ Heisenberg spin-1/2 finite chain,}
Nucl. Phys. B {\bf 554} [FS] (1999),
647--678,
{\tt arXiv:math-ph/9807020}

\bibitem{FS}
O Foda and G Schrader, 
\textit{XXZ scalar products, Miwa variables and discrete KP,}
in 
\textit{New Trends in Quantum Integrable Systems},
B Feigin, M jimbo and M Okado, Editors, 
World Scientific (2010) 61--80,
{\tt  arXiv:1003.2524}

\bibitem{ohta}
Y Ohta, R Hirota, S Tsujimoto and T Inami, 
J. Phys. Soc. of Japan {\bf 62} (1993), 
1872--1886

\bibitem{macdonald.book}
I G Macdonald,
{\it Symmetric functions and Hall polynomials,}
Oxford University Press, (1995)

\bibitem{serban}
D Serban,
{\it A note on the eigenvectors of long-range spin chains 
and their scalar products},
{\tt 1203.5842}

\bibitem{sutherland}
B Sutherland, 
\textit{Low-Lying Eigenstates of the One-Dimensional Heisenberg
Ferromagnet for any Magnetization and Momentum}, 
Phys Rev Lett {\bf 74} (1995) 816.

\bibitem{dhar.shastry}
A Dhar and B Sriram Shastry, 
\textit{Bloch Walls And Macroscopic String States In Bethe's 
Solution Of The Heisenberg Ferromagnetic Linear Chain},
Phys Rev Lett {\bf 85} (2000) 2813.  

\bibitem{caetano.vieira}
J Caetano and P Vieira, private communication.

\bibitem{wheeler.su(3)}
M Wheeler,
{\it Scalar products in generalized models with $SU(3)$-symmetry}, 
{\tt arXiv:1204.2089} 

\bibitem{kuperberg}
G Kuperberg, 
{\it Another proof of the alternating sign matrix conjecture},
International Math Res Notices (1996), No 3, 139--150.

\end{thebibliography}
\end{document}